%% file: main.tex
\documentclass[sigconf]{acmart}

\newcommand{\adi}[1]{\textcolor{black}{#1}}

\usepackage{tabularx}
\usepackage{subcaption}
\usepackage{multirow}

\copyrightyear{2024} 
\acmYear{2024} 
\setcopyright{acmlicensed}\acmConference[UIST '24]{The 37th Annual ACM Symposium on User Interface Software and Technology}{October 13--16, 2024}{Pittsburgh, PA, USA}
\acmBooktitle{The 37th Annual ACM Symposium on User Interface Software and Technology (UIST '24), October 13--16, 2024, Pittsburgh, PA, USA}
\acmDOI{10.1145/3654777.3676392}
\acmISBN{979-8-4007-0628-8/24/10}

\AtBeginDocument{%
  \providecommand\BibTeX{{%
    \normalfont B\kern-0.5em{\scshape i\kern-0.25em b}\kern-0.8em\TeX}}}

\begin{document}
\newcommand{\system}{Augmented Physics}
\title{\system{}: Creating Interactive and Embedded Physics Simulations from Static Textbook Diagrams}

% Creating Embedded Interactive Physics Simulations from Static Textbook Diagrams

% Transforming Static Physics Textbooks into Embedded and Interactive Simulations

% An ML-Integrated Authoring tool for Creating Interactive Physics Simulations from Static Diagrams

% Augmenting Static Physics Textbooks with  Overlaid Interactive Simulations

% Authoring Interactive and Embedded Physics Simulations from Static Textbook Diagrams

\author{Aditya Gunturu}
\affiliation{%
  \institution{University of Calgary}
  \city{Calgary}
  \country{Canada}}  
\email{aditya.gunturu@ucalgary.ca}

\author{Yi Wen}
\affiliation{%
  \institution{Texas A\&M University}
  \city{College Station}
  \country{United States}}
\affiliation{%
  \institution{University of Calgary}
  \city{Calgary}
  \country{Canada}}  
\email{cyberwenyi2357@tamu.edu}

\author{Nandi Zhang}
\affiliation{%
  \institution{University of Calgary}
  \city{Calgary}
  \country{Canada}}  
\email{nandi.zhang@ucalgary.ca}

\author{Jarin Thundathil}
\affiliation{%
  \institution{University of Calgary}
  \city{Calgary}
  \country{Canada}}  
\email{jarin.thundathil@ucalgary.ca}

\author{Rubaiat Habib Kazi}
\affiliation{%
  \institution{Adobe Research}
  \city{Seattle}
  \country{United States}}
\email{rhabib@adobe.com}

\author{Ryo Suzuki}
\affiliation{%
  \institution{University of Colorado Boulder}
  \city{Boulder}
  \country{United States}}
\affiliation{%
  \institution{University of Calgary}
  \city{Calgary}
  \country{Canada}}
\email{ryo.suzuki@colorado.edu}

\renewcommand{\shortauthors}{Gunturu et al.}

\input{0-abstract}

% \begin{teaserfigure}
% \includegraphics[width=0.24\textwidth]{figures/optics1-3.jpg}
% \includegraphics[width=0.24\textwidth]{figures/spring1-2.jpg}
% \includegraphics[width=0.24\textwidth]{figures/pendlum1-3.jpg}
% \includegraphics[width=0.242\textwidth]{figures/circuit3-1.jpg}
% \caption{\system{} is a machine learning-integrated authoring tool to transform static physics diagrams into embedded interactive simulations for various topics, such as optics, kinematics, pendulum, and electric circuits.}
% \label{fig:teaser}
% \end{teaserfigure}

\begin{teaserfigure}
\includegraphics[width=\textwidth]{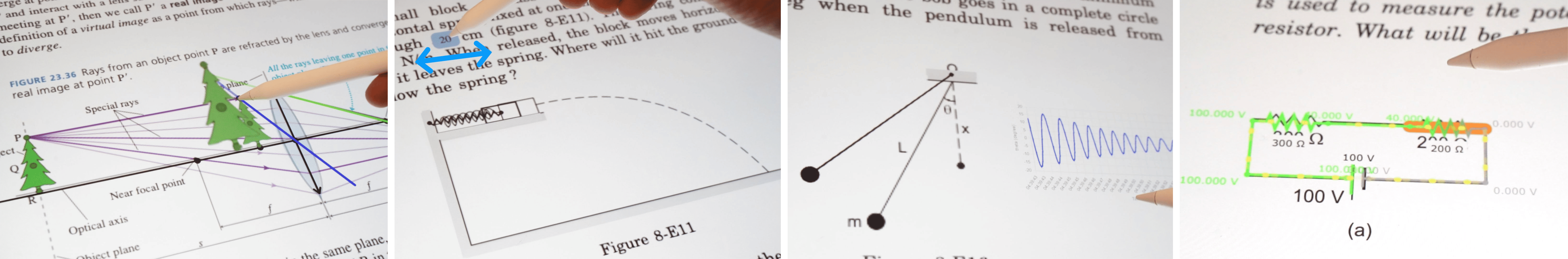}
\caption{\system{} is a machine learning-integrated authoring tool to transform static physics diagrams into embedded interactive simulations for various topics, such as optics, kinematics, pendulum, and electric circuits.}
\Description{Teaser figures with various augmented physics features like augmented experiments and bi-directional manipulatives.}
\label{fig:teaser}
\end{teaserfigure}

\maketitle

\input{1-introduction}
\input{2-related-work}
\input{3-design-space}
\input{4-system}
\input{5-user-study}
\input{6-future-work}

\input{7-conclusion}

\balance

\bibliographystyle{ACM-Reference-Format}
\bibliography{references}

\end{document}

%% file: 0-abstract.tex
\begin{abstract}
We introduce Augmented Physics, a machine learning-integrated authoring tool designed for creating embedded interactive physics simulations from static textbook diagrams. Leveraging recent advancements in computer vision, such as Segment Anything and Multi-modal LLMs, our web-based system enables users to semi-automatically extract diagrams from physics textbooks and generate interactive simulations based on the extracted content. These interactive diagrams are seamlessly integrated into scanned textbook pages, facilitating interactive and personalized learning experiences across various physics concepts, such as optics, circuits, and kinematics. Drawing from an elicitation study with seven physics instructors, we explore four key augmentation strategies: 1) augmented experiments, 2) animated diagrams, 3) bi-directional binding, and 4) parameter visualization. We evaluate our system through technical evaluation, a usability study (N=12), and expert interviews (N=12). Study findings suggest that our system can facilitate more engaging and personalized learning experiences in physics education.
\end{abstract}


\ccsdesc[500]{Human-centered computing~Mixed / augmented reality}
% http://dl.acm.org/ccs.cfm

\keywords{Physics Education; Explorable Explanations; Interactive Paper; Augmented Textbook; Authoring Interfaces}

%% file: 1-introduction.tex
\section{Introduction}
In physics education, interactive simulations play an important role in helping students understand abstract concepts~\cite{perkins2006phet}. Unlike passively reading textbooks, interactive physics simulations enable learners to actively engage with and experiment on complex concepts~\cite{perkins2006phet}. This hands-on approach facilitates a deeper understanding of complex principles~\cite{lee2008explore}, offering a richer and more memorable learning experience than textbooks or videos~\cite{ceberio2016design}.

However, creating these interactive simulations is time-consuming and requires significant effort in programming. 
Given that most educators, as well as students, lack such technical skills, they need to rely on readily available "off-the-shelf" simulations found online.
While these generic simulations can be useful, they sometimes fail to match the exact needs and context of the students' learning materials. For example, our formative study revealed that students frequently struggle to find external simulators that precisely align with textbook content. Moreover, they have to switch back and forth between the website and textbooks to contextualize the online example with their own learning materials. Due to this tedious and distracting process, dynamic simulations are not effectively utilized in current educational settings despite their potential benefits. 

In this paper, we propose \system{}, a novel approach to creating interactive physics simulations by extracting and augmenting content from static textbook diagrams.
By leveraging advanced computer vision techniques like Segment-Anything~\cite{kirillov2023segment} and Multi-modal LLMs, \adi{teachers and educators} can semi-automatically extract diagrams from textbook pages and generate interactive simulations based on the extracted content. 
Our system supports various types of simulations, such as Newtonian motion, optics, circuits, and looping animation (Figure~\ref{fig:teaser}). 
Through a simple authoring process, users can select specific objects in the diagram to segment, manipulate these segmented objects, and adjust parameter values to dynamically interact with simulation results.
Furthermore, these interactive visual outputs are seamlessly overlaid onto the textbook PDF through a web-based interface, allowing students to learn, experiment, and play with their textbooks without needing to search for external materials~\cite{perkins2006phet} or create simulations from scratch~\cite{scott2013physink, cheema2012physicsbook}. 

The idea of creating interactive explanations from static documents is not new~\cite{masson2023charagraph, chulpongsatorn2023augmented}, but this paper contributes in three key ways.
First, we contribute a novel \textit{image-to-simulation} pipeline. Existing works like \textit{Charagraph}~\cite{masson2023charagraph} and \textit{Augmented Math}~\cite{chulpongsatorn2023augmented} mainly focus on \textit{text-to-text} or \textit{text-to-graph} pipelines using standard OCR or simple image \adi{boundary detection}, but such pipelines do not suffice for physics \adi{diagrams and }simulations, which require a more image-centric approach. Therefore, we develop a pipeline to segment diagrams, recognize images, convert them to simulation-ready objects, and integrate them into the textbook diagrams. To the best of our knowledge, our work is the first to explore and demonstrate this image-based physics simulation generation.

Second, we contribute to the design space of augmented physics simulation tools. To design our system, we conducted a formative elicitation study, asking seven physics instructors about how they would augment a physics textbook. Based on the results, we identified four key augmentation strategies: 1) augmented experiments, 2) animated diagrams, 3) bi-directional binding, and 4) parameter visualization.

Third, we contribute insights from three evaluations: a technical evaluation, a preliminary usability study (N=12), and expert interviews with physics instructors (N=12). 

Our technical evaluation results indicate that our pipeline varies based on the types of diagrams, such as kinematics (64\%), optics (44\%), circuits (\adi{40\%}), and animation (66\%), drawn from 200 diagrams across six physics textbooks (these scores reflect the simulation pipeline operating without any modifications, but they are higher when small adjustments are made). Through the user study and expert interviews, we qualitatively compare our approach to existing learning practices, such as handouts, videos, and existing interactive websites, to explore how our tool could meet their needs and fit into current educational practices. Their feedback suggests that our system complements, rather than replaces, existing learning materials such as videos and online simulators. While well-developed existing materials might better work for prepared topics, our tool serves as a way for educators to create on-demand and personalized learning material tailored to specific contexts, which is not well-supported by current practices. Based on their feedback and insights, we discuss ways to expand our proposed approach beyond the current proof-of-concept prototype for future deployment. 

Finally, our main contributions are as follows:
\begin{enumerate}
\item \system{}\footnote{The code and demos are available at \href{https://adigunturu.github.io/AugmentedPhysics/}{\adi{https://adigunturu.github.io/AugmentedPhysics/}}.}, a tool for creating interactive simulations by extracting and animating static physics diagrams.
\item A set of augmentation strategies informed by our formative elicitation study with seven physics instructors.
\item Insights and findings from a technical evaluation, a usability study (N=12), and expert interviews (N=12).
\end{enumerate}

%% file: 2-related-work.tex
\section{Related Work}

\subsection{Physics Simulation for Learning}
Physics simulations have long been recognized as an effective way to enhance learning experiences, particularly within the classroom setting~\cite{kincaid2003simulation, billinghurst2012augmented, garzon2021overview}.
Motivated by this, researchers continuously developed simulated applications~\cite{sari2017effects, ceberio2016design, lee2008explore, kaufmann2008simulating} to help students explore complex physics concepts and foster a deeper understanding. 
For example, various online physics simulators such as \textit{PhET}~\cite{phet}, \textit{MyPhysicsLab}~\cite{myphysicslab}, \textit{the Physics Classroom}~\cite{physicsclassroom}, \textit{oPhysics}~\cite{oPhysics}, \textit{Physion}~\cite{physion}, and \textit{Simphy}~\cite{Simphy} facilitate understanding of various physics concepts such as kinematics, magnetism, sound, and circuits.
Beyond screen-based physics simulations, HCI researchers have also explored AR and tangible physics simulation tools (e.g., \textit{Bogusevschi et al.}~\cite{bogusevschi2020teaching}
\textit{Cai et al.}~\cite{cai2013using}, \textit{Thees et al.}~\cite{thees2020effects} \textit{Radu et al.}~\cite{radu2019can}, \textit{RealitySketch}~\cite{suzuki2020realitysketch}, \textit{ConductAR}~\cite{narumi2015conductar}, \textit{Urp}~\cite{underkoffler1999urp}, \textit{HOBIT}~\cite{furio2017hobit}, \textit{Illuminating Clay}~\cite{piper2002illuminating}, \textit{Physics Playground}~\cite{kaufmann2008simulating}, \textit{Sketched Reality}~\cite{kaimoto2022sketched}, \textit{Physica}~\cite{li2023physica}, \textit{CircuitTUI}~\cite{CircuitTUI}), which provide more engaging and collaborative experiences through spatial and embodied interactions.

However, these existing physics simulation tools are often limited to pre-programmed and off-the-shelf simulations, which sometimes fail to meet the specific needs and challenges students face. To address this limitation, HCI researchers have explored authoring tools that allow users to create personalized physics simulations on demand.
Tools like \textit{PhysInk}~\cite{scott2013physink}, \textit{PhysicsBook}~\cite{cheema2012physicsbook}, \textit{MathPad2}~\cite{LaViola2007mathPad2}, and \textit{ChalkTalk}~\cite{perlin2018chalktalk} enable users to sketch physics diagrams, which are then automatically transformed into interactive and animated graphics. Such authoring tools allow non-technical users like students and instructors to easily and quickly create physics simulations without programming skills. 
\adi{While these tools hold great potential, they do not focus on simulating existing figures but on creating them from scratch, which can become tedious. Textbooks are primary teaching materials and contain rich and expressive diagrams for various concepts, which can be used to create personalized and situated learning experiences for students. Moreover, extracting textbook content also allows teachers to create higher-quality simulations that are identical and situated to student's textbooks (which they can take home) compared to low-fidelity sketches. Our formative study reveals that students often need clear guidance and instructions to find related resources.} To fill this gap, this paper explores an alternative approach to generating interactive simulations by animating existing static diagrams instead of sketching them from scratch.

\subsection{Augmenting Existing Documents}
Previous research has investigated how to make static explanations more dynamic and interactive.
For instance, Victor introduced the concept of \textit{Explorable Explanations}~\cite{victor2011explorable}, demonstrating various interactive explanations for scientific reading~\cite{victor2005magic, victor2011killmath, victor2013media}. 
Such interactive explanations enhance readers' understanding by highlighting the relationship between texts and data~\cite{latif2021kori, kong2014extracting}, allowing in-situ exploration through multiverse analysis~\cite{dragicevic2019increasing} among others~\cite{hohman2020communicating}.
However, a key limitation of interactive documents is the inherent need for programming, which requires substantial time and cost to create them~\cite{head2022math}.
While tools like \textit{Tangle}~\cite{victor2013tangle}, \textit{Idyll Studio}~\cite{conlen2021idyll}, and \textit{Data Theater}~\cite{lau2020data} aim to lower this barrier, they still need programming, leaving existing static documents unusable for interactive explanations. 

To address this problem, researchers have developed methods to augment existing documents, rather than programming them from scratch.
Prior works have investigated tools to semi-automatically generate summaries (e.g., \textit{Marvista}~\cite{chen2022marvista}), references (e.g., \textit{HoloDoc}~\cite{Li2019holodoc}), highlights (e.g., \textit{ScentHighlights}~\cite{chi2005scenthighlights}, \textit{Scim}~\cite{fok2023scim}, \textit{Kim et al.}~\cite{kim2018facilitating}), annotations (e.g., \textit{Threddy}~\cite{kang2022threddy}, 
\textit{Contextifier}~\cite{hullman2013contextifier}, \textit{textSketch}~\cite{subramonyam2020texsketch}, \textit{DuallyNoted}~\cite{qian2022dually}), 
and visualizations (e.g., \textit{Elastic Documents}~\cite{badam2018elastic}, \textit{Jessica et al.}~\cite{hullman2018improving}) by augmenting existing documents.
Most closely related to our work, \textit{Charagraph}~\cite{masson2023charagraph} and \textit{Augmented Math}~\cite{chulpongsatorn2023augmented} explore the semi-automatic generation of interactive charts and graphs by extracting text from static documents.
 % However, these text-based extraction methods fall short for physics simulations, which heavily rely on visual representation, rather than textual input. 
\adi{While our work shares a similar motivation with some others: to make textbooks interactive and explorable, our goals and methods differ fundamentally. While other works focus on augmenting text content, our system focuses on making diagrams themselves explorable and animated by extracting and simulating individual components. This is crucial for Physics education, where diagrams represent dynamic systems and processes that change over time. High school physics concepts rely on visualizing motion, and our augmentation strategies enable users to animate individual components in the image and make them interact with each other to craft engaging and explorable experiences.}

% Thus, our paper contributes a new pipeline to generate interactive \adi{simulations} directly from images, by leveraging object segmentation, image recognition, and physics simulator generation. 
% To the best of our knowledge, our approach of image-based extraction represents the first attempt at generating interactive explanations directly from images, rather than from text.

\subsection{Tools for Authoring Interactive Diagrams}
Previous research has explored end-user authoring tools for creating dynamic and interactive diagrams for various applications, including technical illustrations~\cite{zhu2011sketch}, scientific explanation~\cite{sarracino2017user}, and artistic animation~\cite{xing2016energy}. In the educational domain, many online tools~\cite{GeoGebra95:online,oPhysics} and research prototypes~\cite{saquib2021constructing, LaViola2007mathPad2} allow for interactive authoring and animation. 
These tools enable users to create animation through sketch-based interactions~\cite{willett2018mixed, davis2008k, kazi2014draco, kazi2016motion, kazi2014kitty} and tangible demonstrations~\cite{barnes2008video}.
Such authoring techniques have been demonstrated to be versatile and adaptable across various domains, including 3D animations~\cite{ma2022layered}, video augmentation~\cite{hashim2023drawing, xia2023realitycanvas}, and motion graphics videos~\cite{jahanlou2022katika}. Although these methods have significantly enhanced the creative authoring of dynamic visuals, they may not be ideally tailored for explanatory content in physics textbooks. In these contexts, animated objects must adhere to specific physical behaviors, making the general techniques potentially less effective.

Similar to our focus, several tools have been developed to animate static documents.
For example, \textit{Revision}~\cite{savva2011revision} helps users bind corresponding data with text in the document, and \textit{PaperTrail}~\cite{rajaram2022paper} augments static documents through manual demonstration.
Building upon these works, we have recognized the immense potential of interactive visuals for educational purposes. Our objective is to broaden this scope, enabling both educators and learners to effortlessly create their interactive diagrams within textbook pages, which facilitates a richer learning experience through intricate explorations.

%% file: 3-design-space.tex
\section{Formative Study}
To design our system, we conducted a formative study with seven physics instructors. The goals of this formative study were twofold: 1) to understand their current methods of teaching and learning physics to identify gaps and needs in current educational practices, and 2) to gather insights into potential augmentation strategies through design elicitation, guiding the design of such a tool from a pedagogical perspective.

\subsection{Method}
\subsubsection{Participants}
We recruited seven participants from our local university community (6 males, 1 female). The participants, all students with substantial backgrounds in physics education, represented the full spectrum of educational attainment in physics, including undergraduate (1), master's (5), and PhD candidates (1) from the physics department and related disciplines. On average, participants had 1.7 years of teaching experience as TAs or instructors. Each study session lasted approximately one hour, and in exchange for their time, all participants received a \$10 Amazon gift card upon completion of the study.

\subsubsection{Protocol}
After obtaining consent, we provided participants with a primer on HCI research and described the goals of our exploration and the formative study. First, we conducted an open-ended discussion with participants to explore their views on current instructional practices in physics, identifying pedagogical gaps and needs for a potential new tool.

Second, we conducted a design elicitation study to speculate on a new tool to fill these gaps. For the elicitation study, each participant was provided with the same textbook: \textit{``Physics for Scientists and Engineers: A Strategic Approach, 3rd Edition" by Randall D. Knight''}~\cite{knight2017physics}, a typical first-year physics textbook for undergraduate students. We chose this textbook because it includes many diagrams that participants could use and covers a wide range of topics in physics, including kinematics, circular motion, Newtonian mechanics, electromagnetism, light and optics, and circuits.

As participants explored the text, we asked them to imagine how the static concept diagrams they encountered could be augmented to enhance their understanding of the underlying concepts. Participants were also instructed to approach the task from the perspective of a teacher. We asked participants to elicit possible designs using a think-aloud protocol. Additionally, we provided them with stationery to mark up the textbook with illustrations, which we later translated into figures in subsequent sections.

\subsection{Challenges of Current Practices}
The results of our formative study highlight several pedagogical limitations of current practices in physics education and the clear need for augmented interactive explanations to bridge these educational gaps.

\subsubsection{Static Visualizations Cannot Represent Time-Dependent Physics Concepts}
Most educational materials for physics currently rely on static visualizations. Participants mentioned that static visualizations tend to suffice when illustrating simple concepts, allowing students to grasp the underlying principles. However, these static visualizations become non-intuitive when depicting concepts involving motion or systems that change over time. For instance, P2 referred to a diagram of gravitational potential energy and expressed a wish to \textit{``point at this object and see what forces it is undergoing at this point in time''}. Regarding time-dependent behavior, P4 pointed out that \textit{``the behavior of bodies in an elliptical orbit was not accurately illustrated by a static diagram''}, failing to illustrate the varying velocity of a celestial object as it progresses through its orbit. 

\subsubsection{Videos Enhance Understanding but Lack Experimentation Opportunities}
Participants noted that undergraduate physics students are often directed to watch YouTube videos on a topic to gain a better understanding of concepts that are difficult to grasp through static visualizations. However, these videos, as per the participants, also come with their limitations in terms of interaction and the ability to experiment. For example, P1 mentioned that \textit{``YouTube videos are not interactive''}, and that \textit{``being able to interact helps you with building intuition''}. The absence of interactivity was seen as a drawback because intuitive learning in physics is heavily reliant on experimentation.

\subsubsection{Simulation Tools Lack Sufficient Instructional Scaffolding}
Most participants were familiar with online simulation tools but noted that these simulators often require students to create their physics simulations, assuming a solid understanding of the subject. Using a circuit simulator as an example, they highlighted that students are expected to build circuits from scratch. While this promotes open-ended experimentation, it can leave students uncertain about how to begin. In contrast, textbooks offer scaffolding through existing diagrams, potentially reducing the steps needed to create a meaningful simulation. Textbooks, according to P2 and P7, already provide \textit{``guiding steps''} that aid in understanding a topic.
Thus, participants felt that while simulation tools are beneficial adjuncts to other materials, relying exclusively on them can pose challenges.

Turning to external resources to supplement classroom physics teaching introduces two significant challenges for students: the content may not directly align with the classroom's unique curriculum, and deviating from primary materials can lead to distractions. By enhancing the static diagrams already present in classroom resources, \system{} directly addresses these issues. Several participants recognized the benefit of improving visuals from their study materials over seeking external sources. They believed that examples introduced in the classroom provide a fundamental understanding, which could be further enriched by additional augmented visuals.

\subsubsection{External Content Might Misalign and Distract from Core Learning}
Beyond these drawbacks, seeking external content to supplement classroom materials presents two significant challenges. First, external content may not always align closely with the concepts as taught in the classroom. Given that existing simulation tools offer generalized experiments, students must manually contextualize and bridge the gap between them.
Second, by shifting focus away from core materials, students often face distractions, such as other content or recommendations in YouTube videos.
In light of these issues, several participants emphasized the clear advantage and necessity of augmenting visualizations found in their materials rather than sourcing them externally.
They believed that classroom-introduced examples provided a foundational mental scaffolding, which could be enhanced further with augmented visualizations.

\subsection{Elicited Augmentation Strategies}
In the development of our system, we collected design suggestions from participants on a broad array of topics, including kinematics, optics, electromagnetism, Newtonian gravity, acoustics, and thermodynamics. From their feedback, we identified four primary categories of augmentation techniques. This section outlines these techniques, supplemented by sketches that illustrate the participants' ideas.

\subsubsection*{\textbf{Augmented Experiments}}
The most popular approach was to dynamically simulate diagrams based on physics principles, allowing students to interact with concepts depicted in textbooks and visualize experiments through real-time feedback. For example, participants envisioned observing the path of light rays bending when the position of a lens was altered. They highlighted the importance of such simulations for gaining an intuitive understanding of underlying concepts. Additionally, participants expressed the desire to modify simulation parameters, such as altering the lens's refractive index to observe its impact on light rays. This desire extended to parameters like mass or velocity in collision simulations, emphasizing the need for simulations to respond to user-defined changes. For instance, if a diagram illustrated two orbiting bodies and their masses were altered, the simulation should adjust the orbit’s barycenter and eccentricity accordingly.

\begin{figure}[h!]
\centering
\includegraphics[width=\linewidth]{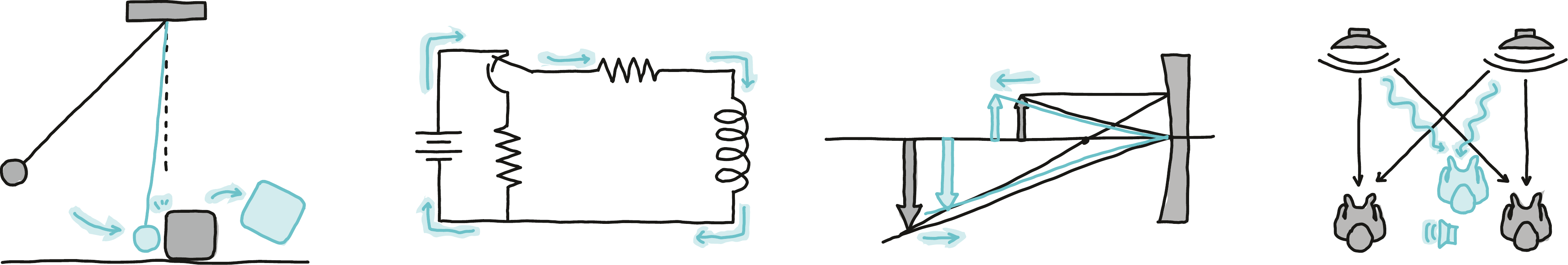}
\caption { Augmented Experiments:  \adi{Enabling users to directly manipulate textbook diagrams, enabling them to change parameters such as the position of an object in an optics diagram or the resistance in a circuit diagram, and observe real-time changes.}}
\Description{4 sketches depicting various examples of Augmented Experiments.}
\label{fig:augmented-experiments}
\end{figure}

\subsubsection*{\textbf{Animated Diagrams}}
Animating static diagrams emerged as another primary technique from participant feedback. In contrast to augmented experiments, this technique focuses more on repeated animation rather than simulated behaviors. Participants unanimously expressed the wish to see diagrams dynamically demonstrating changes over time, as static diagrams often fail to convey evolving systems adequately. In the acoustics domain, for instance, there was a noted need for animations that depict the continuous movement of sound and electromagnetic waves to foster better understanding. Recognizing the educational value of animations seen in YouTube videos, participants believed that even simple animations, such as an object tracing an orbital path, could significantly enhance the intuitiveness and engagement of concepts.

\begin{figure}[h!]
\centering
\includegraphics[width=\linewidth]{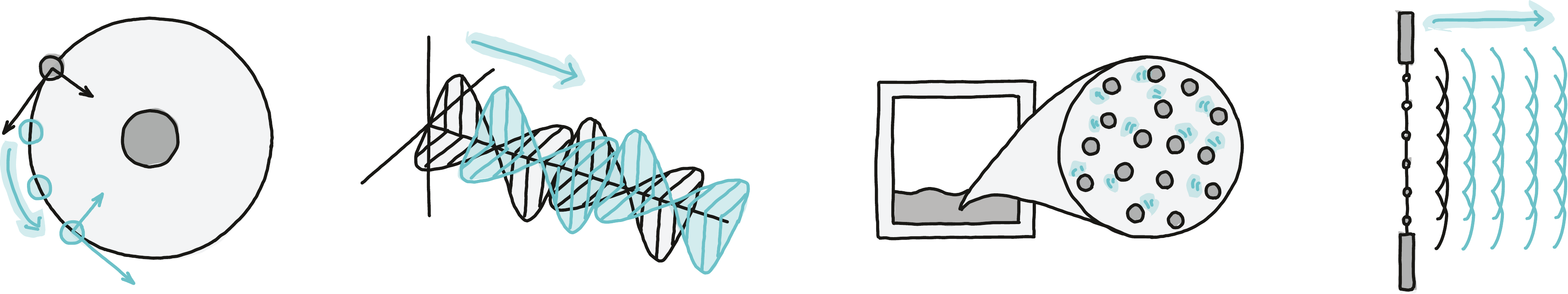}
\caption{Animated Diagrams: \adi{Converting static figures into looped dynamic animations, showing changes over time.}}
\Description{4 sketches depicting various examples of Animated Diagrams.}
\label{fig:animated-diagrams}
\end{figure}

\subsubsection*{\textbf{Bi-Directional Binding}}
Physics often involves bi-directional relationships between measurable parameters of a system. Examples in kinematics include the relationships between kinetic and potential energies or between mass and acceleration. Textbooks typically present these concepts alongside data visualizations and graphical depictions of real-world scenarios. An example provided in our study was a diagram showing a ball dropped from a certain height, accompanied by a bar chart of the changing kinetic and potential energy as the ball falls. Participants indicated that the static nature of these visualizations limited their understanding. Instead of static data visualizations, they preferred the option to adjust the ball's height on the page and observe accurate reflections of these changes in the bar chart. Further discussion revealed a desire for the reverse: manipulating the bar chart to see the ball's height adjust. Thus, participants were interested in operationalizing the bi-directional relationships between system parameters through the augmentation of static diagrams.

\begin{figure}[h!]
\centering
\includegraphics[width=\linewidth]{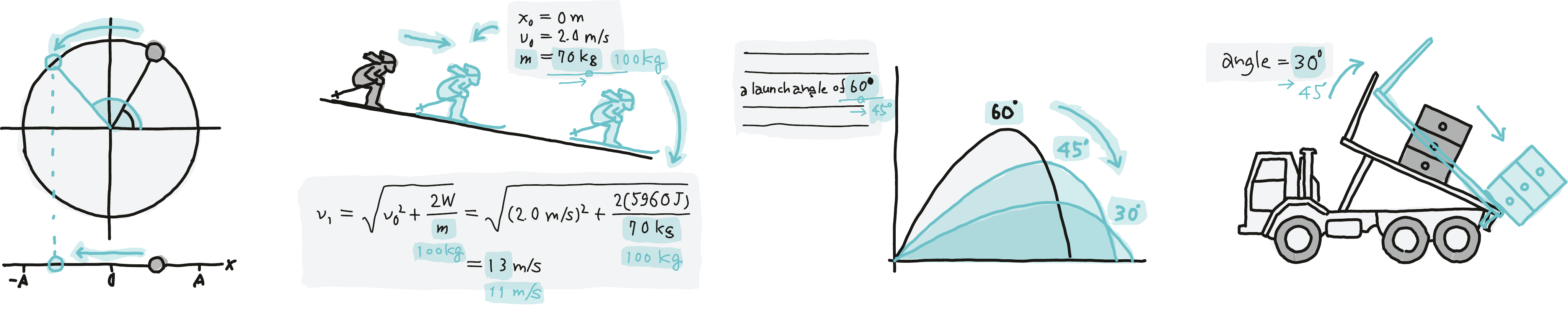}
\caption{Bi-Directional \adi{Binding: Connecting text to the diagrams and making them manipulable.}}
\Description{4 sketches depicting various examples of Bi-Directional Binding.}
\label{fig:bi-directional-parameters}
\end{figure}

\subsubsection*{\textbf{Parameter Visualization}}
Participants expressed interest in creating data visualizations for diagrams in the textbook that lacked accompanying visual data. For example, concerning circuits, two participants mentioned the potential benefits of a digital oscilloscope that could measure voltage across any two points on a circuit diagram. Additionally, one participant proposed representing the motion of a planet in an elliptical orbit on a velocity-time graph, showcasing the planet's increased velocity at perihelion with changes in orbital eccentricity.

\begin{figure}[h!]
\centering
\includegraphics[width=\linewidth]{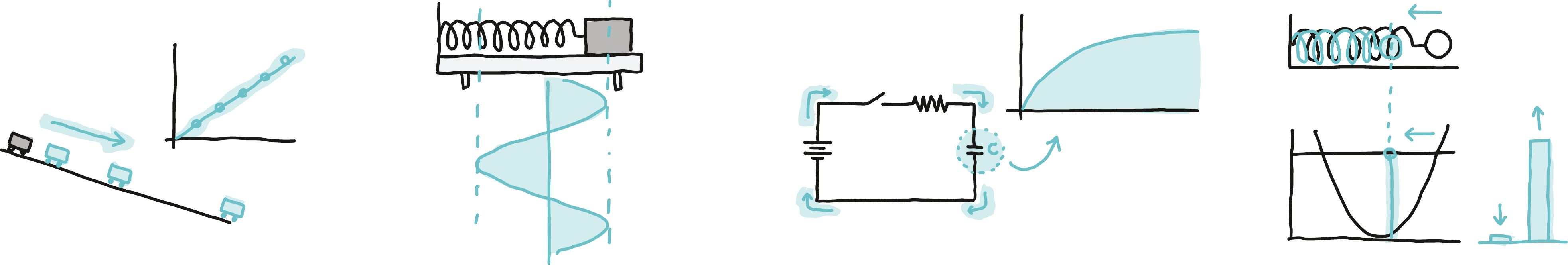}
\caption{Parameter Visualization: \adi{Generating on-demand visualizations of various parameters in the simulated diagram.}}
\Description{4 sketches depicting various examples of Parameter Visualization.}
\label{fig:parameter-visualization}
\end{figure}

%% file: 4-system.tex
\section{\system{}: System Design}

\subsection{Overview}
In this section, we introduce \system{}, a machine learning-integrated authoring tool designed to enable non-technical users to create interactive physics simulations from static diagrams. Our web-based tool facilitates users, including students and instructors, in semi-automatically extracting diagrams from physics textbooks and generating simulations that seamlessly integrate with scanned textbook pages. Our research primarily focuses on basic physics concepts taught in high schools across the United States, such as Newtonian motion, optics, and electric circuits. Although more advanced topics like quantum mechanics are beyond our current scope, our adaptable animated diagrams technique allows users to create animated illustrations for these concepts as well. Moreover, we have made our system open-source, including the machine learning pipeline and browser-based simulators, to encourage the HCI community to further develop our prototype and methods.

\begin{figure}[h!]
\centering
\includegraphics[width=0.32\linewidth]{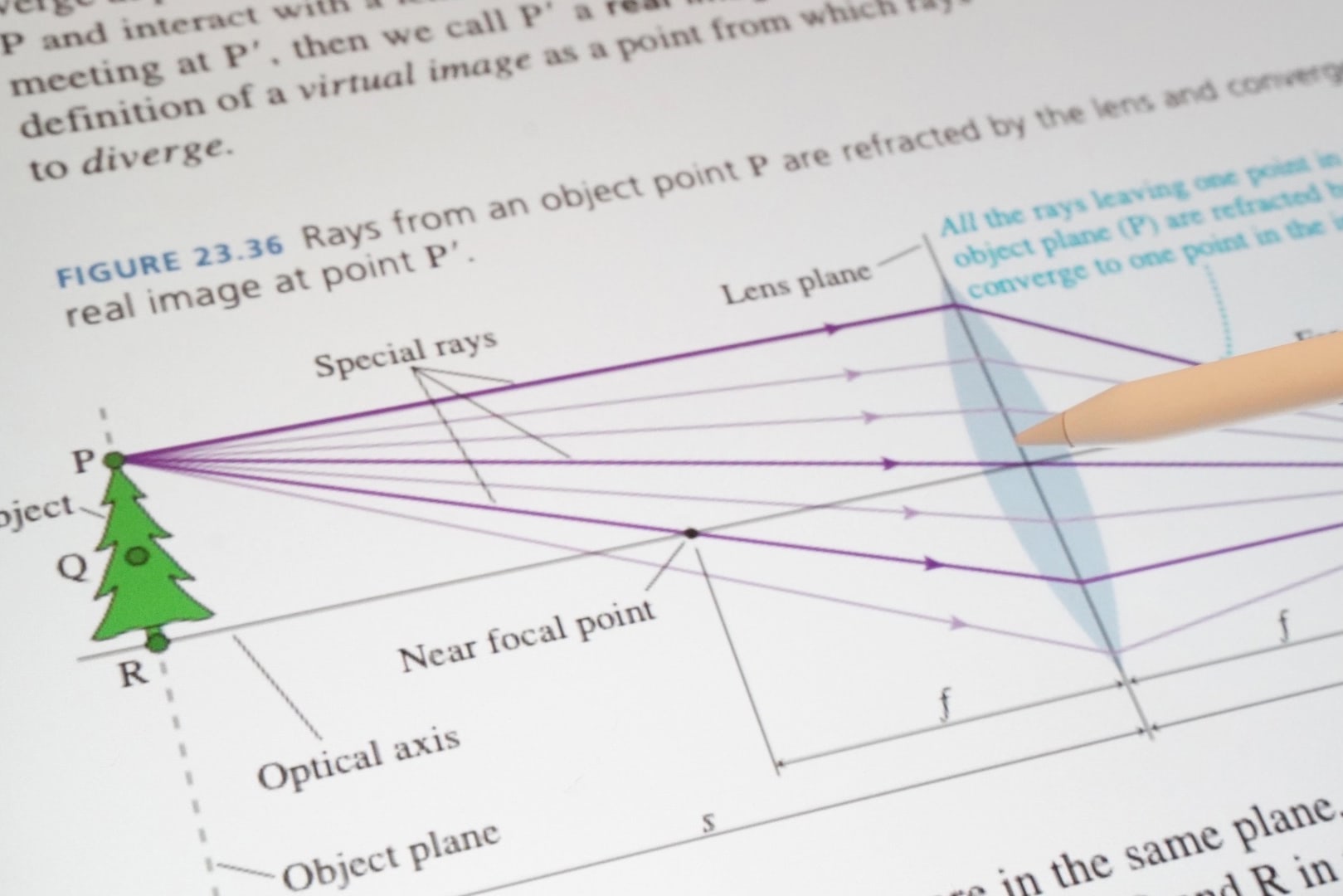}
\includegraphics[width=0.32\linewidth]{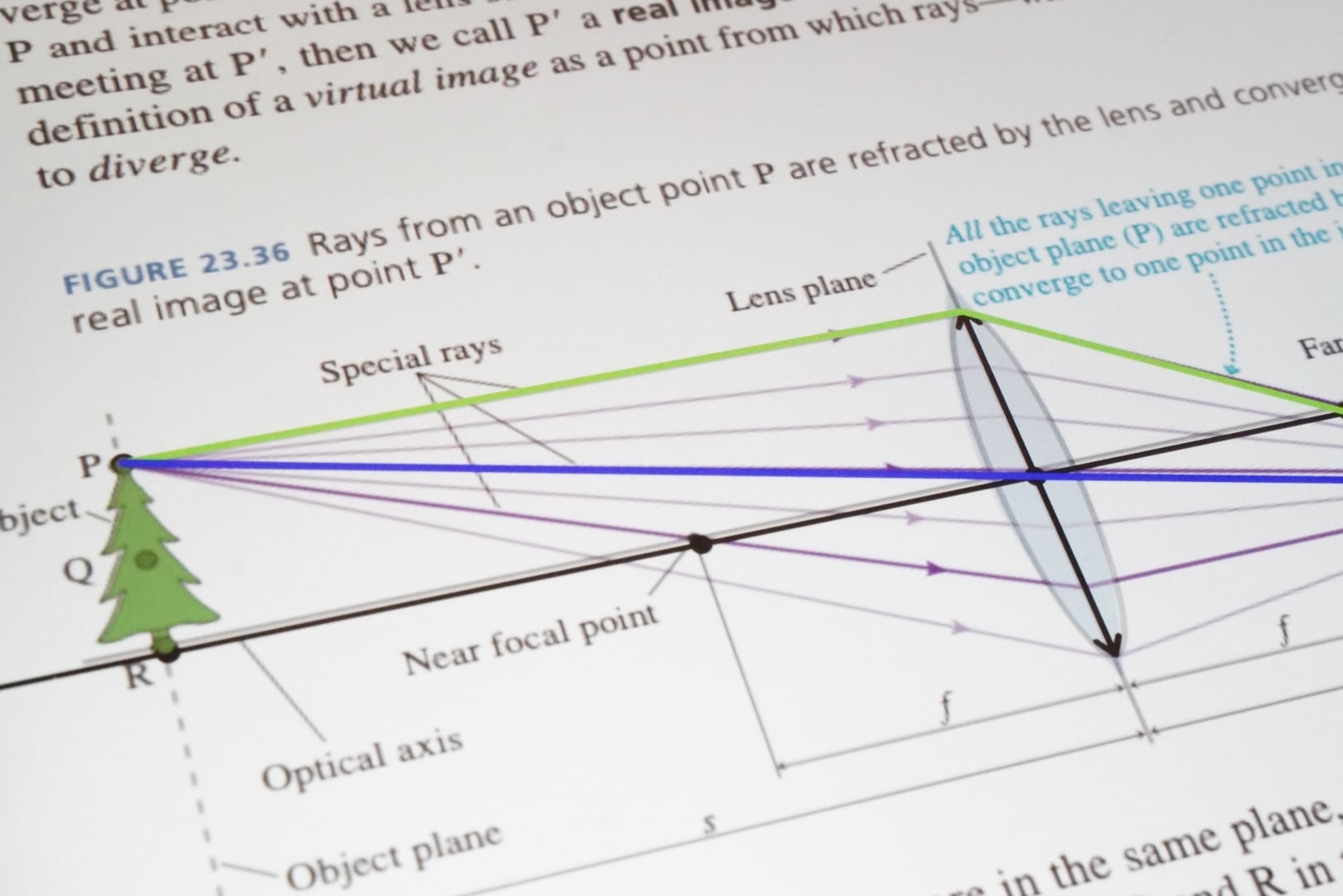}
\includegraphics[width=0.32\linewidth]{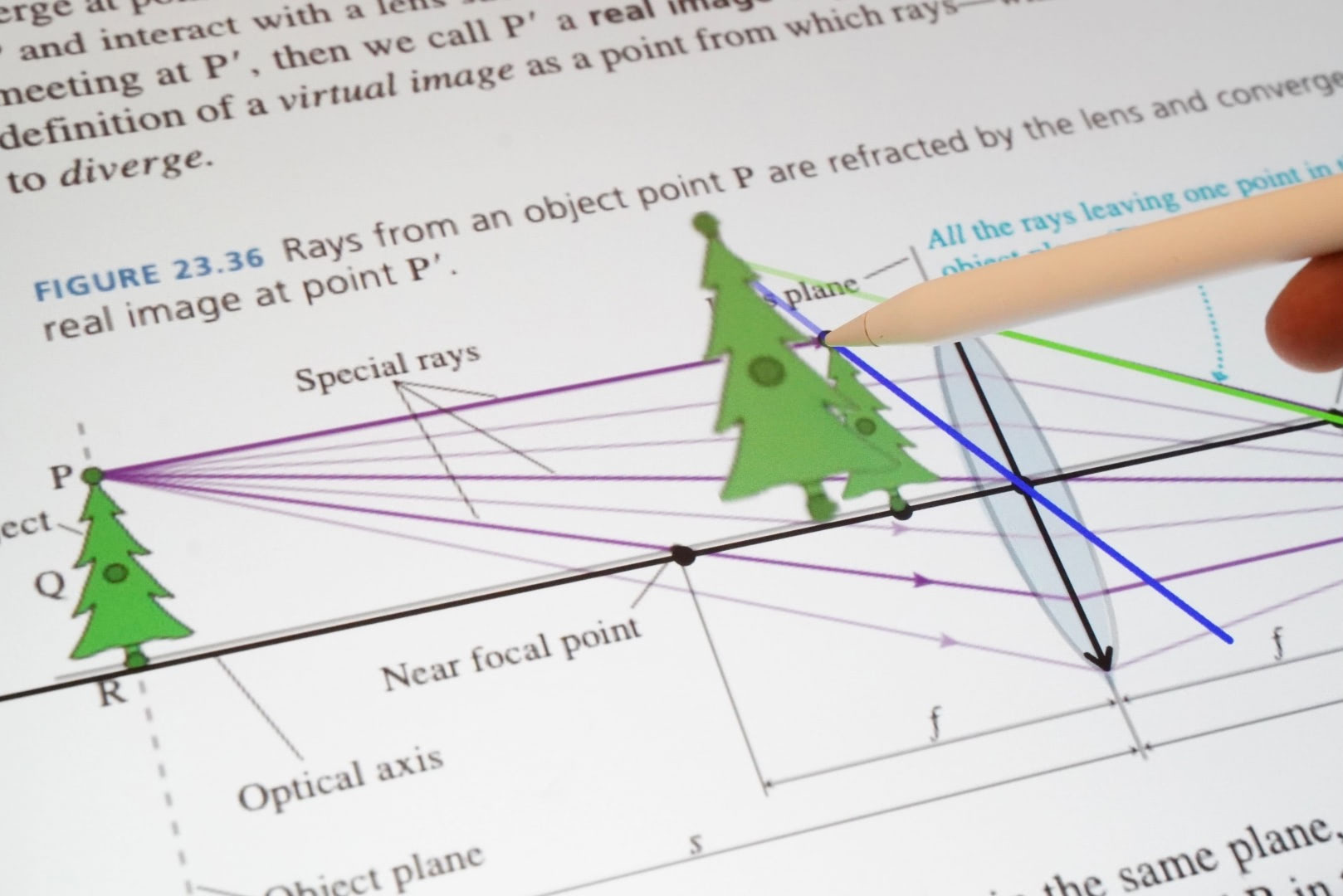}
\caption{Interactive simulations for an optics diagram. 1) The user segments objects, lenses, and focal points. 2) The system generates an overlaid simulation. 3) The user interacts with the object and focal point to observe changes.}
\label{fig:optics1}
\Description{3 images of an optics diagram being simulated with interactivity.}
\end{figure}

\subsection{Authoring Workflow}
Our authoring workflow includes the following steps: 1) Import a textbook page, 2) Choose a simulation type, 3) Extract and segment images, 4) Assign roles to the segmented image, 5) Generate and run a simulation, and 6) Interact with simulation results through parameter manipulation.
In the following sections, we illustrate this workflow using a series of examples drawn from high school physics textbooks.

\subsubsection*{\textbf{Step 1. Import a Textbook Diagram}}
The initial step involves the user importing a diagram through our web interface. Our system supports both desktop and mobile devices, allowing users to either upload a PDF of a textbook page from their computer or capture and upload a picture of the textbook page using their smartphone.

\subsubsection*{\textbf{Step 2. Choose a Simulation Type}}
Upon importing the diagram, the system asks the user to select a type of simulation from the available options. \adi{The system first automatically recommends a simulation type.} Users have the choice among three specific simulation categories: \textit{kinematics}, \textit{optics}, and \textit{circuits}. There is also an option for \textit{animation}, catering to scenarios that do not necessitate a particular type of simulation.

\subsubsection*{\textbf{Step 3. Segment Images}}
\begin{figure}[h!]
\centering
\includegraphics[width=0.32\linewidth]{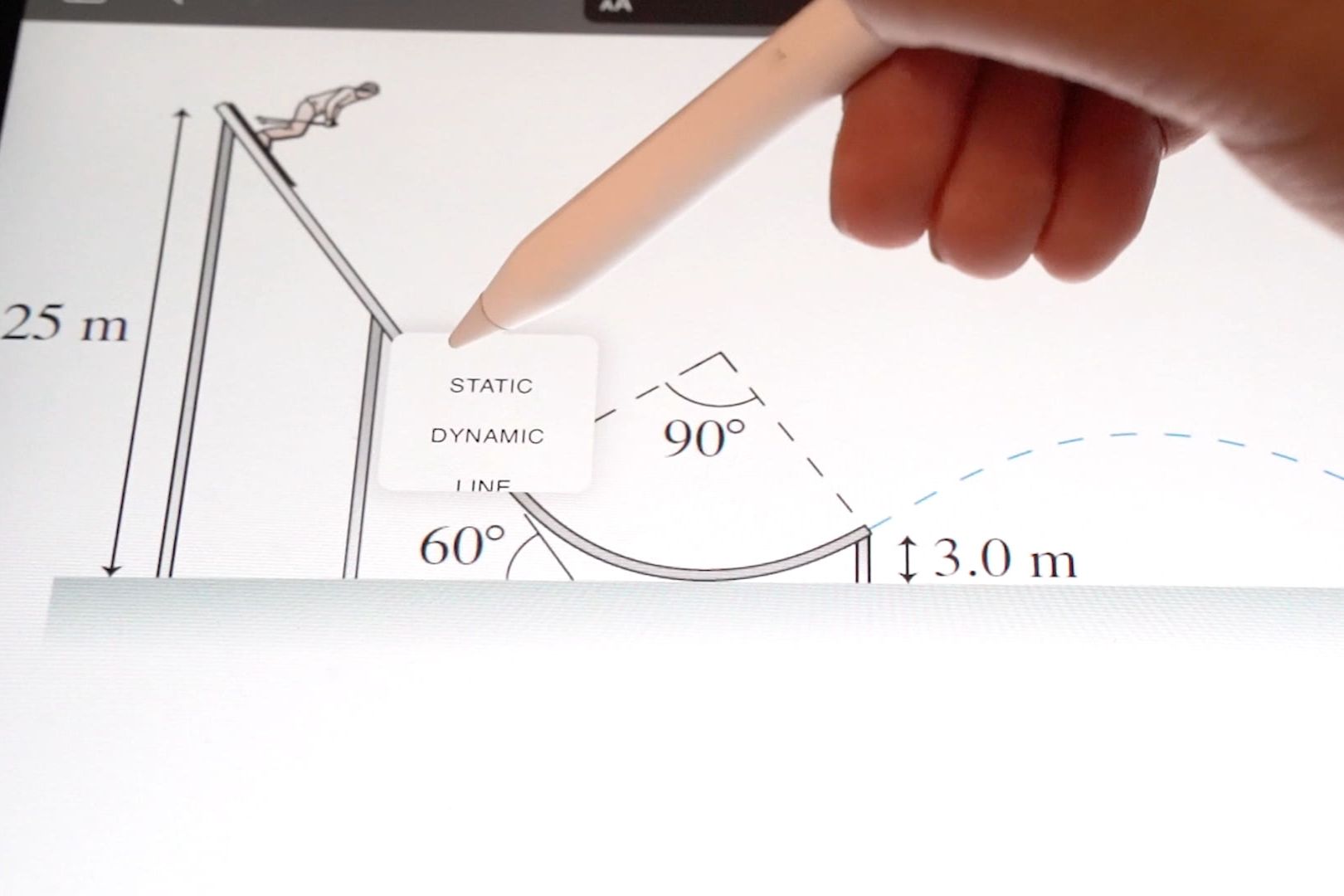}
\includegraphics[width=0.32\linewidth]{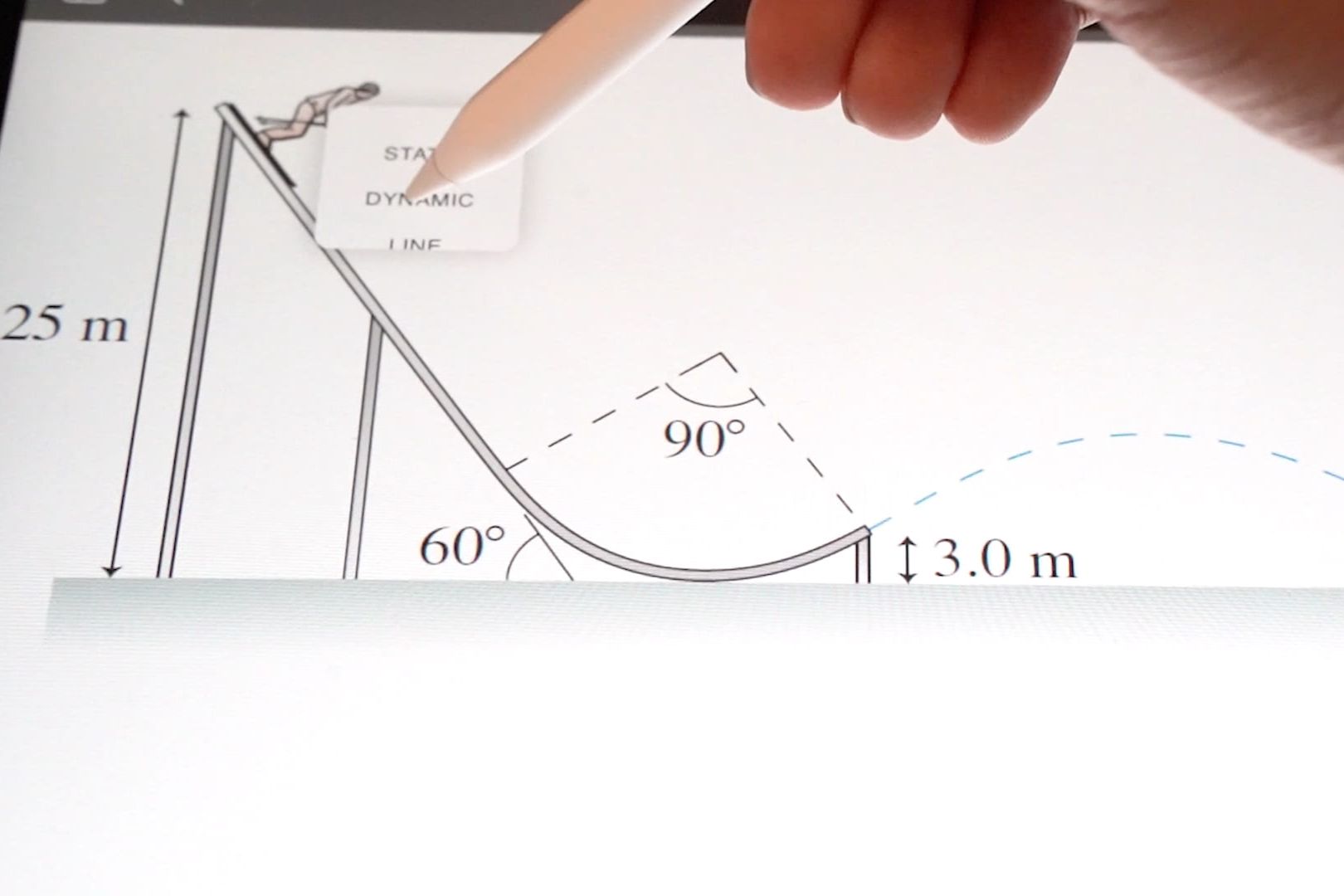}
\includegraphics[width=0.32\linewidth]{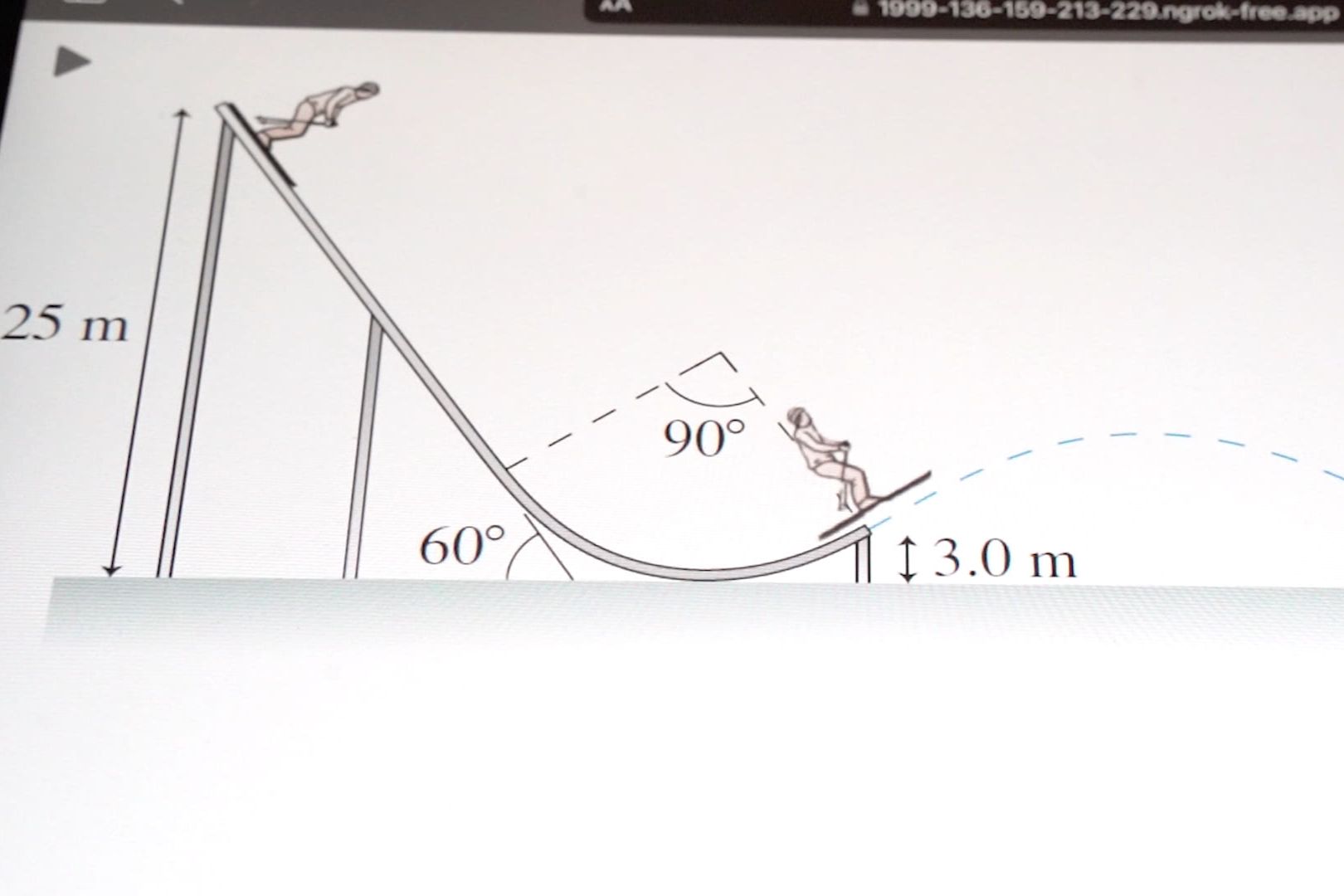}
\caption{Interactive simulations for a kinematics diagram. 1) The user segments a slope as a static object. 2) The user segments a skier as a dynamic object. 3) The system generates and runs a simulation.}
\label{fig:gravity1}
\Description{3 images of a kinematic diagram being simulated as an augmented experiment.}
\end{figure}

The next step involves image segmentation. The user initiates segmentation by selecting a specific area on the diagram with either a box or a point. For instance, a user might select a tree and a lens to segment these objects from an optics-related diagram (Figure~\ref{fig:optics1}). In another case, users can segment various objects, such as objects and slopes, in a diagram related to Newtonian motion (Figure~\ref{fig:gravity1}). Additionally, users can also segment a line to extract a path for creating a line-following animation (Figure~\ref{fig:animation2}).

\subsubsection*{\textbf{Step 4. Assign Roles to Segmented Objects}}
Following segmentation, the user assigns a label to each segmented object, providing them with specific roles within the simulation. The system presents a set of available roles for each simulation type, allowing the user to select from them. For example, in an optics simulation, a user might label a tree as an \textit{object to be projected}, a lens simply as \textit{lens}, and a point as the \textit{focal point} (Figure~\ref{fig:optics1}). In gravity-related scenarios, segmented items can be classified as \textit{dynamic objects}, affected by gravity, or \textit{static objects}, which remain stationary (Figure~\ref{fig:gravity1}). Additional labels such as \textit{spring} or \textit{string} can be used for kinematics diagrams (Figure~\ref{fig:spring1}). For circuit simulations, the system automatically classifies objects, such as \textit{resistors} and \textit{batteries}, using image recognition (Figure~\ref{fig:circuit1}). 

\subsubsection*{\textbf{Step 5. Generate and Run a Simulation}}
\begin{figure}[h!]
\centering
\includegraphics[width=0.32\linewidth]{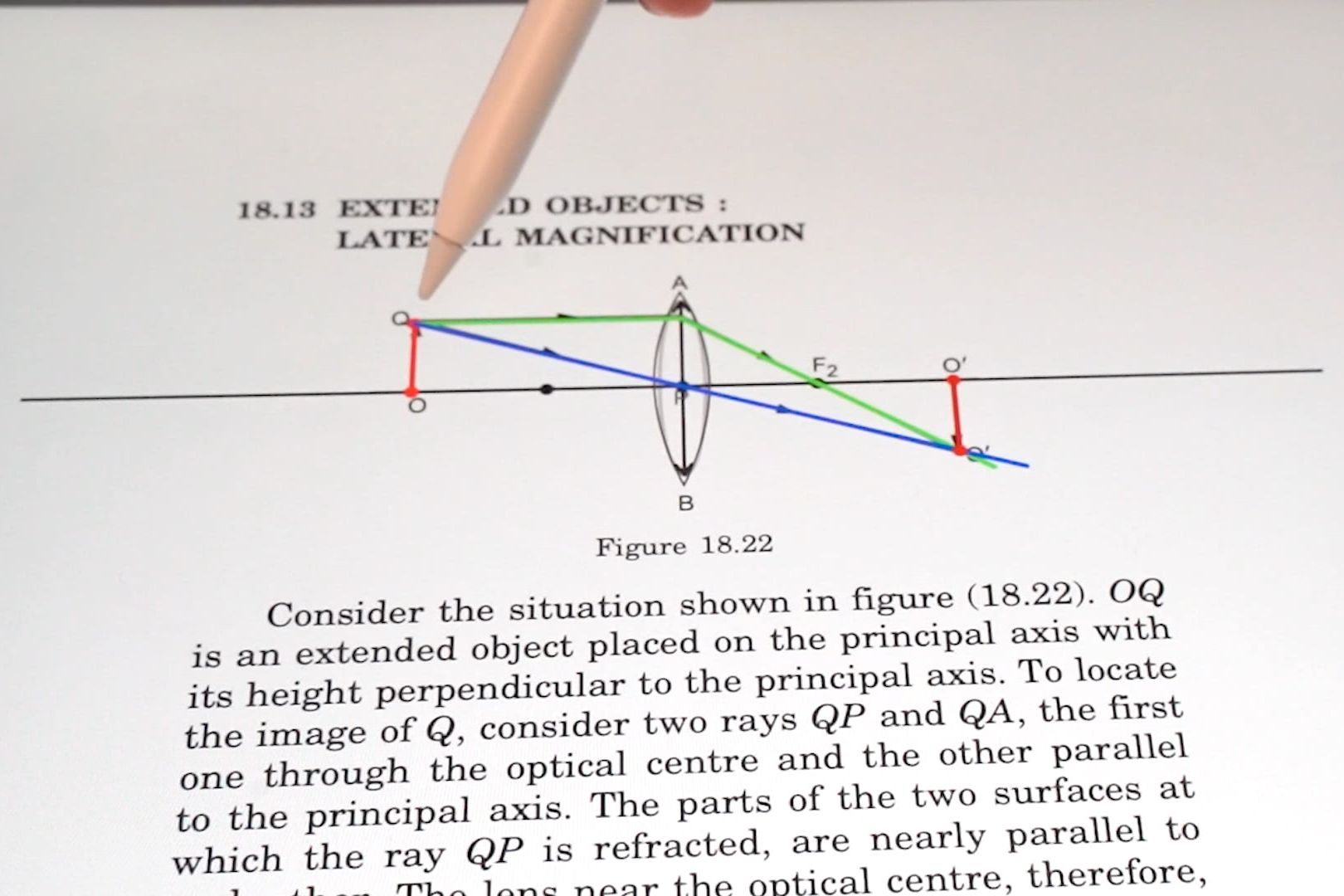}
\includegraphics[width=0.32\linewidth]{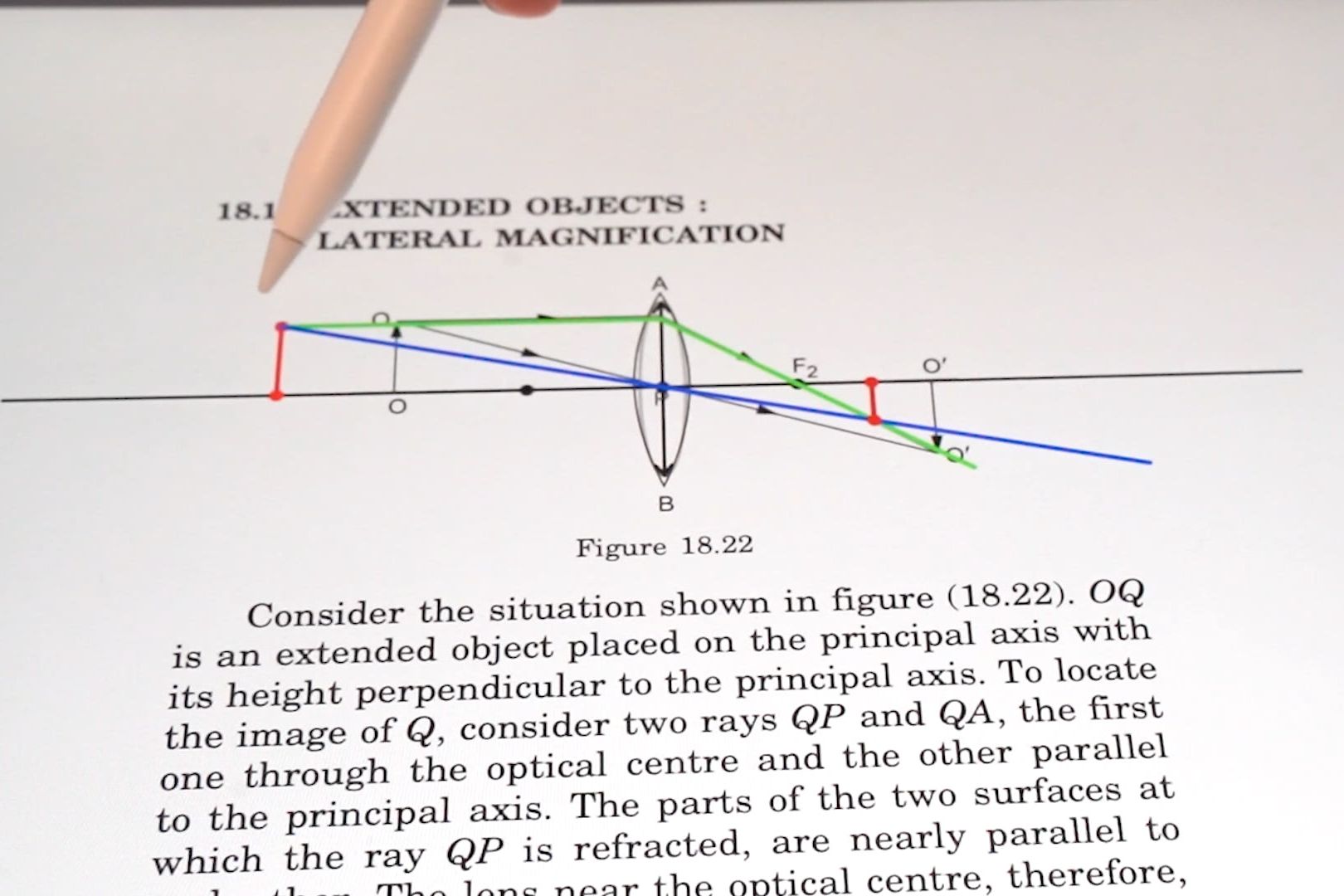}
\includegraphics[width=0.32\linewidth]{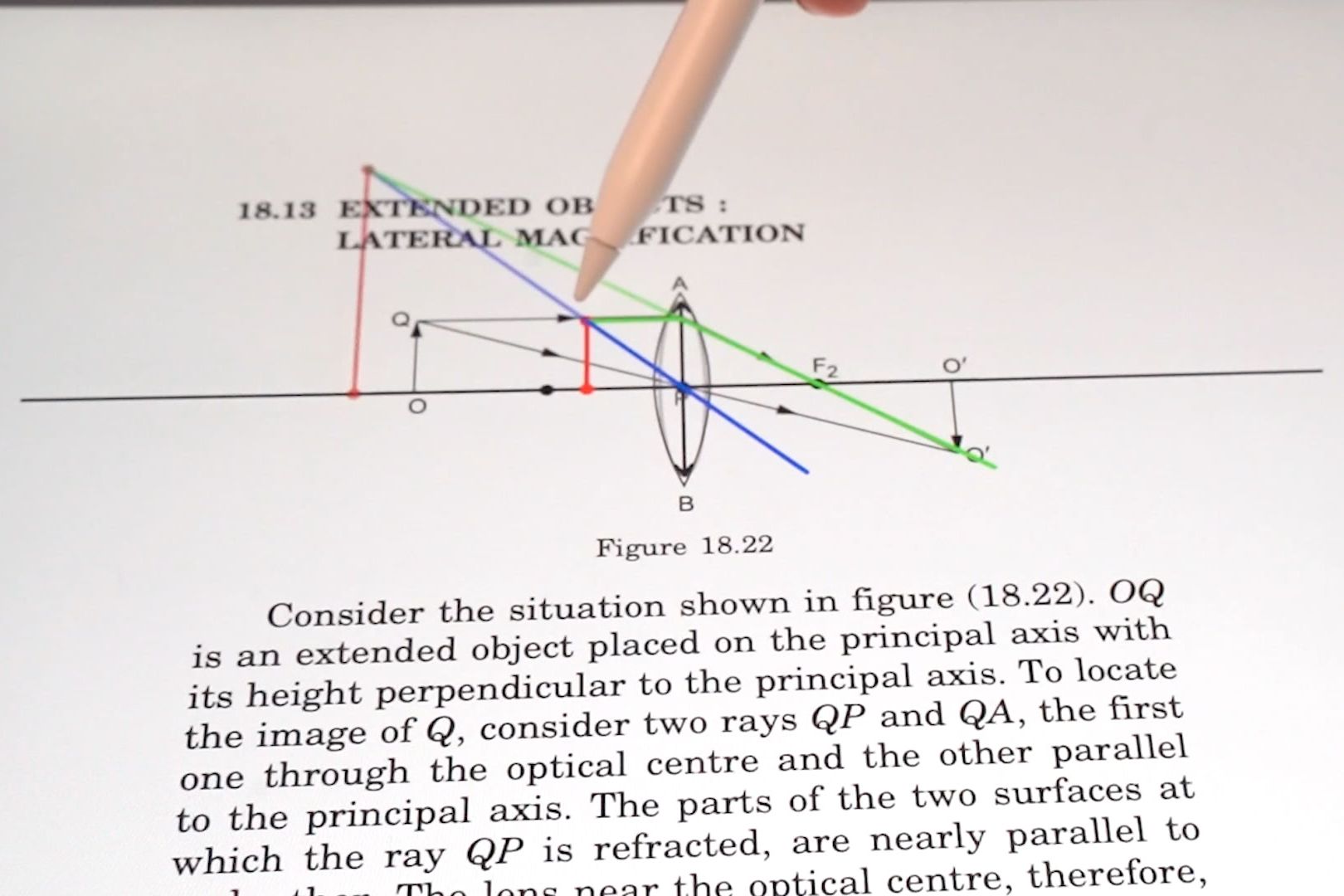}
\caption{Interactive simulations for a different optics diagram, where the user can see the interactive simulation result.}
\label{fig:optics2}
\Description{Another 3 images of an optics diagram being simulated with interactivity.}
\end{figure}

Once users have segmented images and assigned roles, the system proceeds to generate the simulation by converting the segmented images into polygons with appropriate properties for the physics simulation. For example, the skier and slope shown in Figure~\ref{fig:gravity1} are precisely replicated to create polygons for dynamic and static objects, respectively. This approach ensures the simulation integrates seamlessly with the original diagram, achieving alignment in both shape and position within the image. After completing these steps, the simulation is ready to be launched. Users can start it by clicking the run button \adi{Figure} ~\ref{fig:gravity1} or by interacting with the rendered polygons to witness dynamic visual outputs (Figure \ref{fig:optics2}). They can click on the simulated objects and optionally change parameters.

\subsubsection*{\textbf{Step 6: Interact with the Simulation through Parameter Manipulation}}
Users have the flexibility to adjust parameters within the simulation. Depending on their roles, different objects come with various parameters, such as mass for dynamic objects, friction for static objects, and force constants for springs. Moreover, the system can recognize parameter values within text or images, enabling users to manipulate numerical values on the page. For example, in electrical circuit simulations, users can modify values like those of resistors and batteries to dynamically change the simulation results. \adi{Additionally, the system automatically links numerical values from the text to specific properties of objects in the simulation, which the user can edit.}

\subsection{Supported Augmentation Features}
We have developed the following four augmentation techniques: 1) \textit{augmented experiments}, 2) \textit{animated diagrams}, 3) \textit{bi-directional binding}, and 4) \textit{parameter visualization}. These features are tailored to support a wide range of simulated experiments that educators wish to create for their students.

\subsubsection{\textbf{Augmented Experiments}}
\begin{figure}[h!]
\centering
\includegraphics[width=0.32\linewidth]{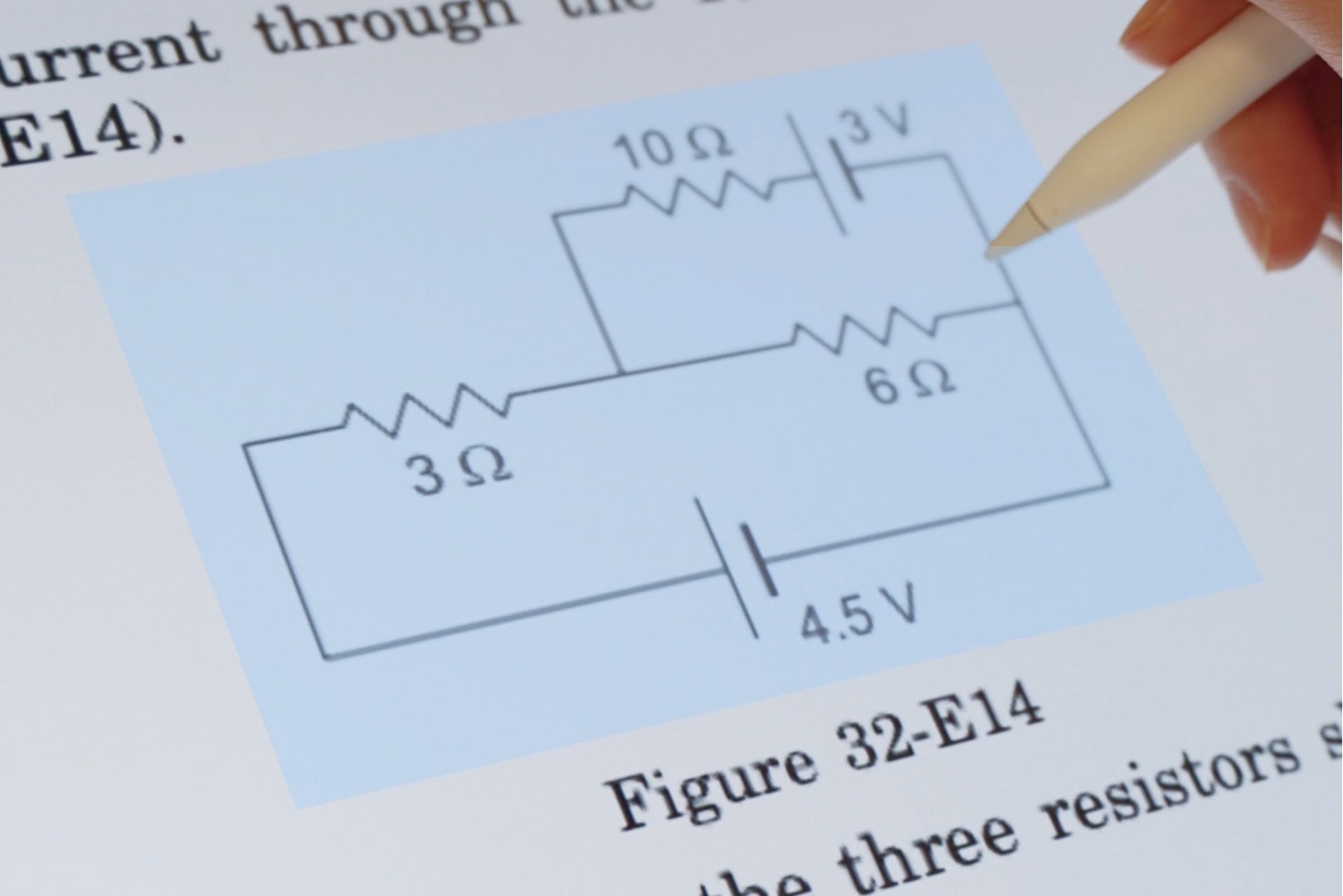}
\includegraphics[width=0.32\linewidth]{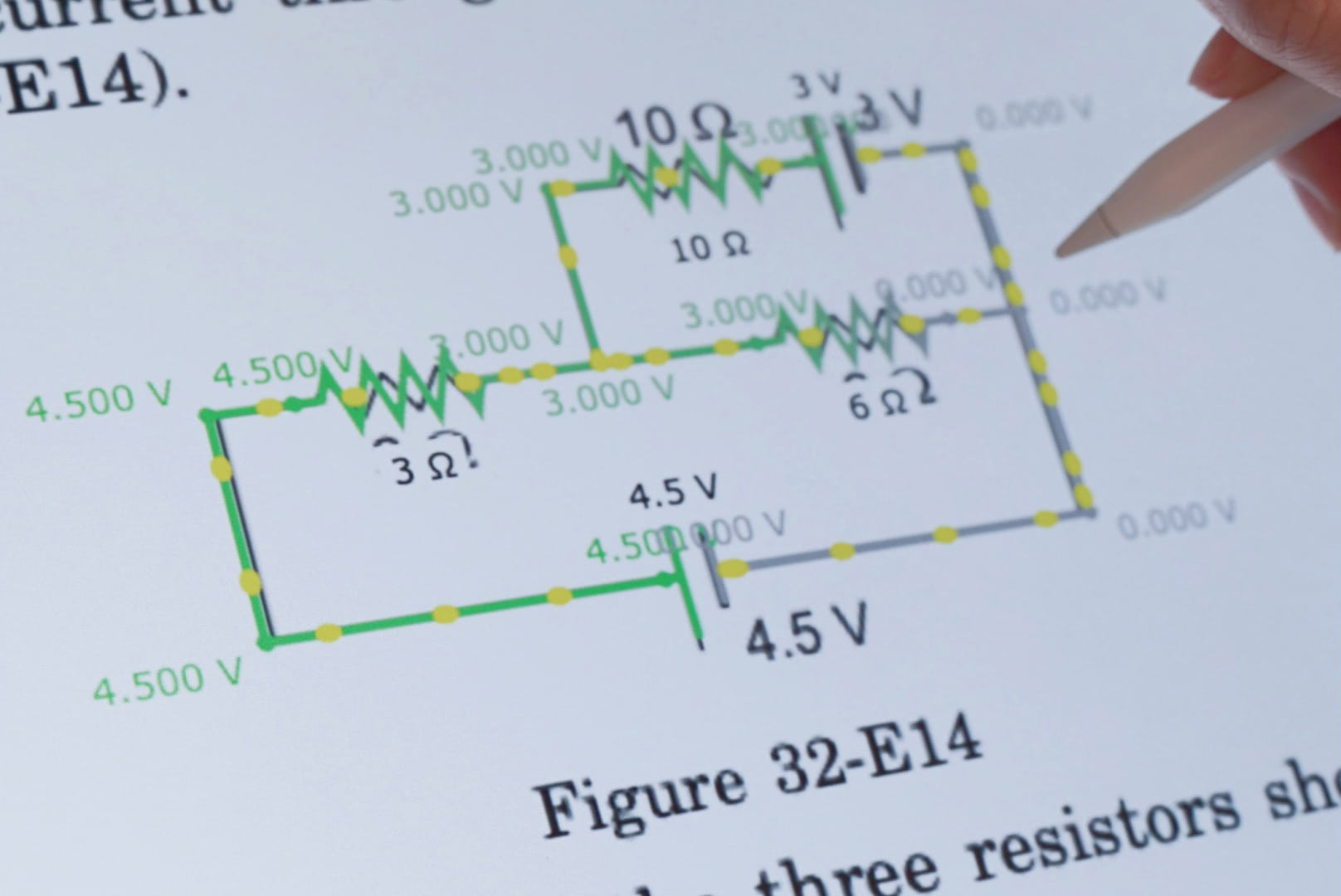}
\includegraphics[width=0.32\linewidth]{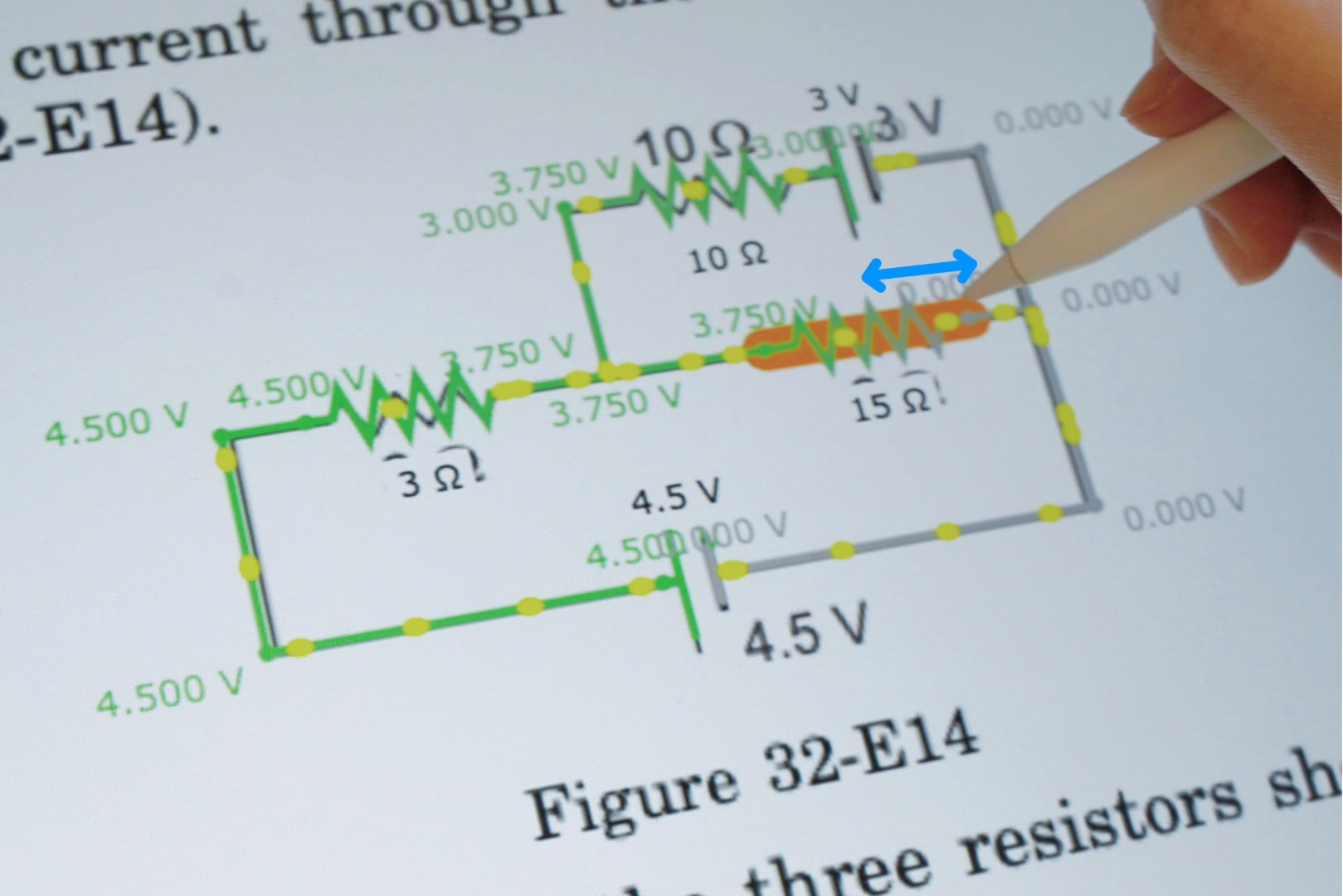}
\caption{Augmented experiments for an electrical circuit diagram. 1) The user first selects a diagram. 2) The system then generates an overlaid simulation. 3) The user interacts with the simulation's values.}
\label{fig:circuit1}
\Description{Images of circuits being simulated with animations.}
\end{figure}

Augmented experiments transform textbook images into interactive simulations, enabling students to manipulate parameters and interact with the diagrams. For instance, as illustrated in Figure~\ref{fig:optics1}, students can drag a tree object closer to a convex lens within the simulation to observe the formation of a virtual image on the same side as the object. Alternatively, in circuit simulations as shown in Figure~\ref{fig:circuit1}, users can modify the voltage and register value of each electronic component, which in turn alters the current flow. This allows them to observe real-time changes in amperage and voltage across points within the circuits. 

As previously stated, our system supports three simulation categories: 1) kinematics, covering gravity, Newtonian motion, springs, and pendulums; 2) optics, focusing on lenses, light propagation, and image formation through mirroring; and 3) electric circuits, dedicated to simulating current flow in electronic circuits.

\subsubsection{\textbf{Animated Diagrams}}
\begin{figure}[h!]
\centering
\includegraphics[width=0.32\linewidth]{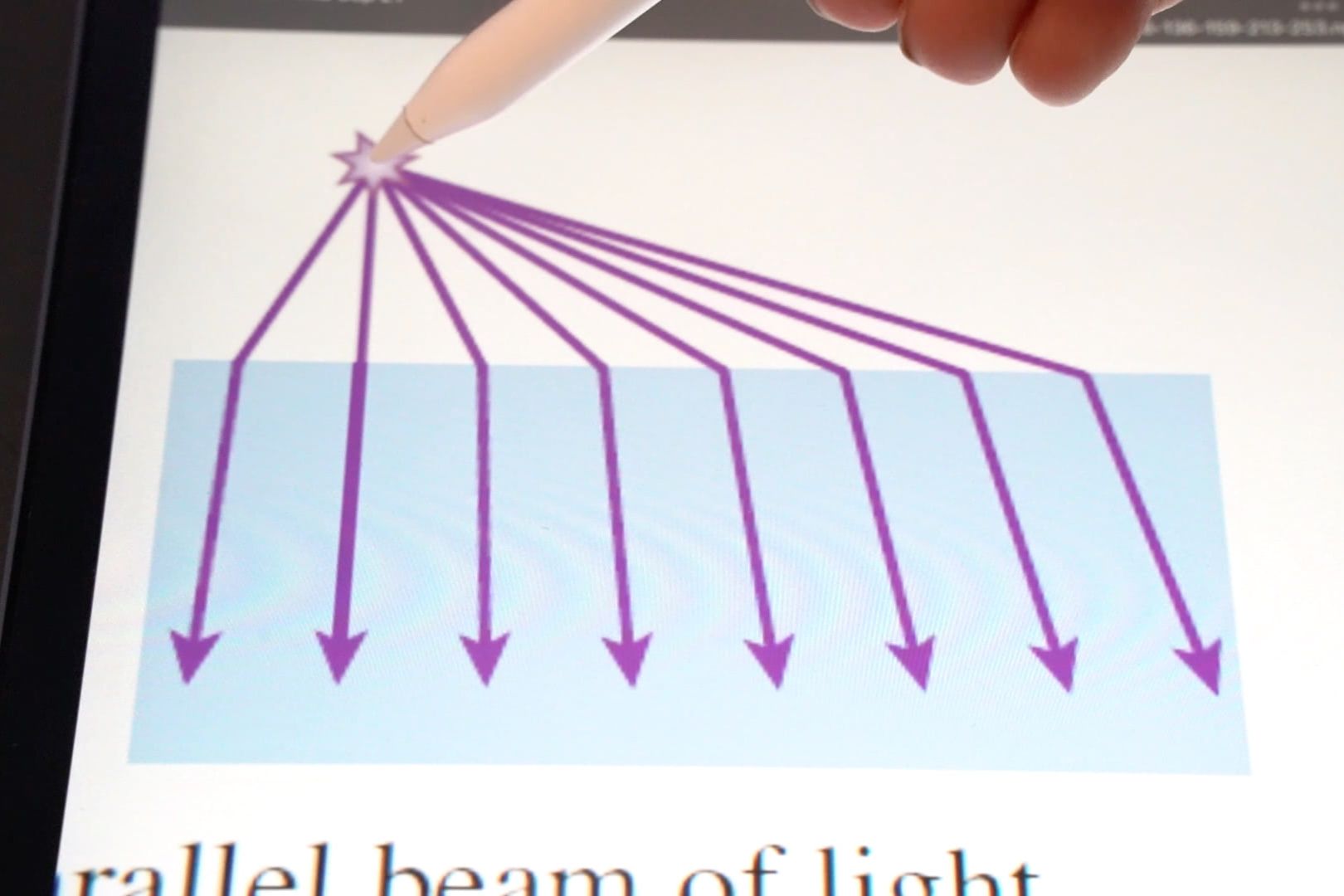}
\includegraphics[width=0.32\linewidth]{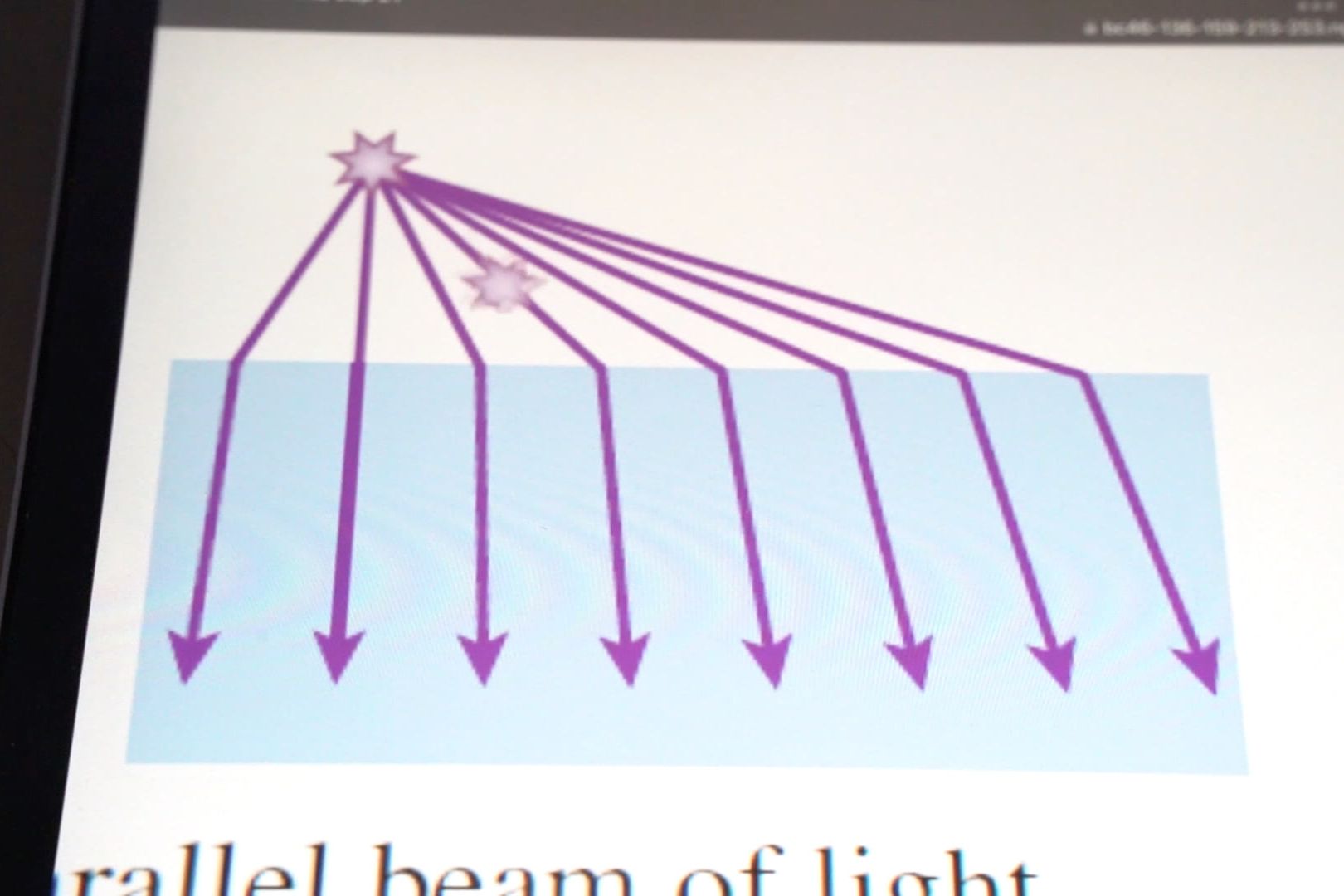}
\includegraphics[width=0.32\linewidth]{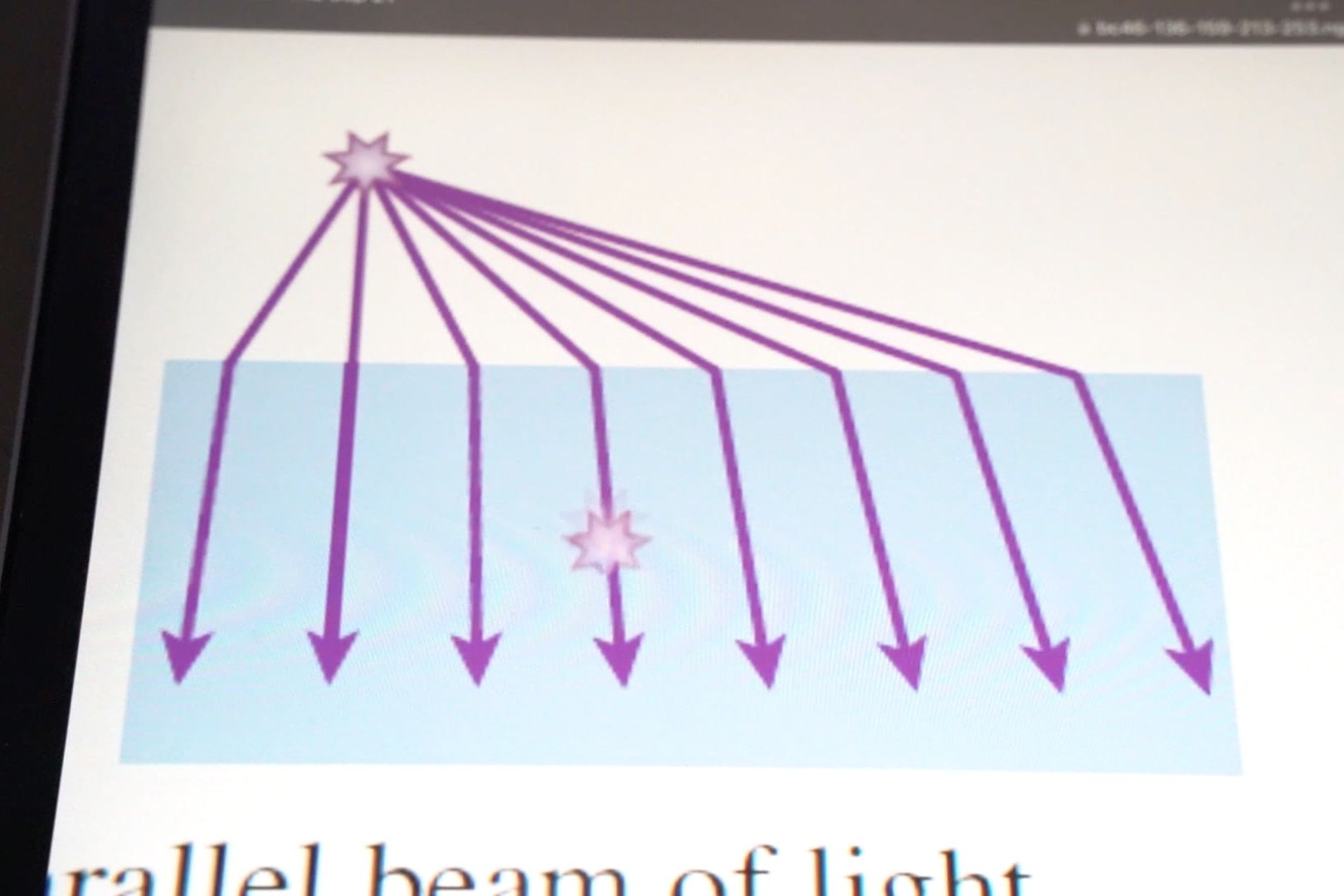}
\caption{Animated diagrams for a light refraction diagram. 1) The user segments the photon. 2) The user segments a refraction path. 3) The system animates the photon through the path.}
\label{fig:animation2}
\Description{Images of animated diagrams in a physics textbook.}
\end{figure}

Animated diagrams offer a method to create recurring animations. Users can designate paths for segmented objects to follow, thus creating animations that simulate movements. For example, Figure \ref{fig:animation2} demonstrates how light follows various paths of reflection based on the angle, achieved by segmenting the object and defining a path for the animation. This feature facilitates the creation of captivating animations directly from textbook content, such as the Earth orbiting the Sun. Furthermore, unlike augmented experiments, which are limited to available simulations, animated diagrams can be applied to any diagram.

\subsubsection{\textbf{Bi-directional Binding}}
\begin{figure}[h!]
\centering
\includegraphics[width=0.32\linewidth]{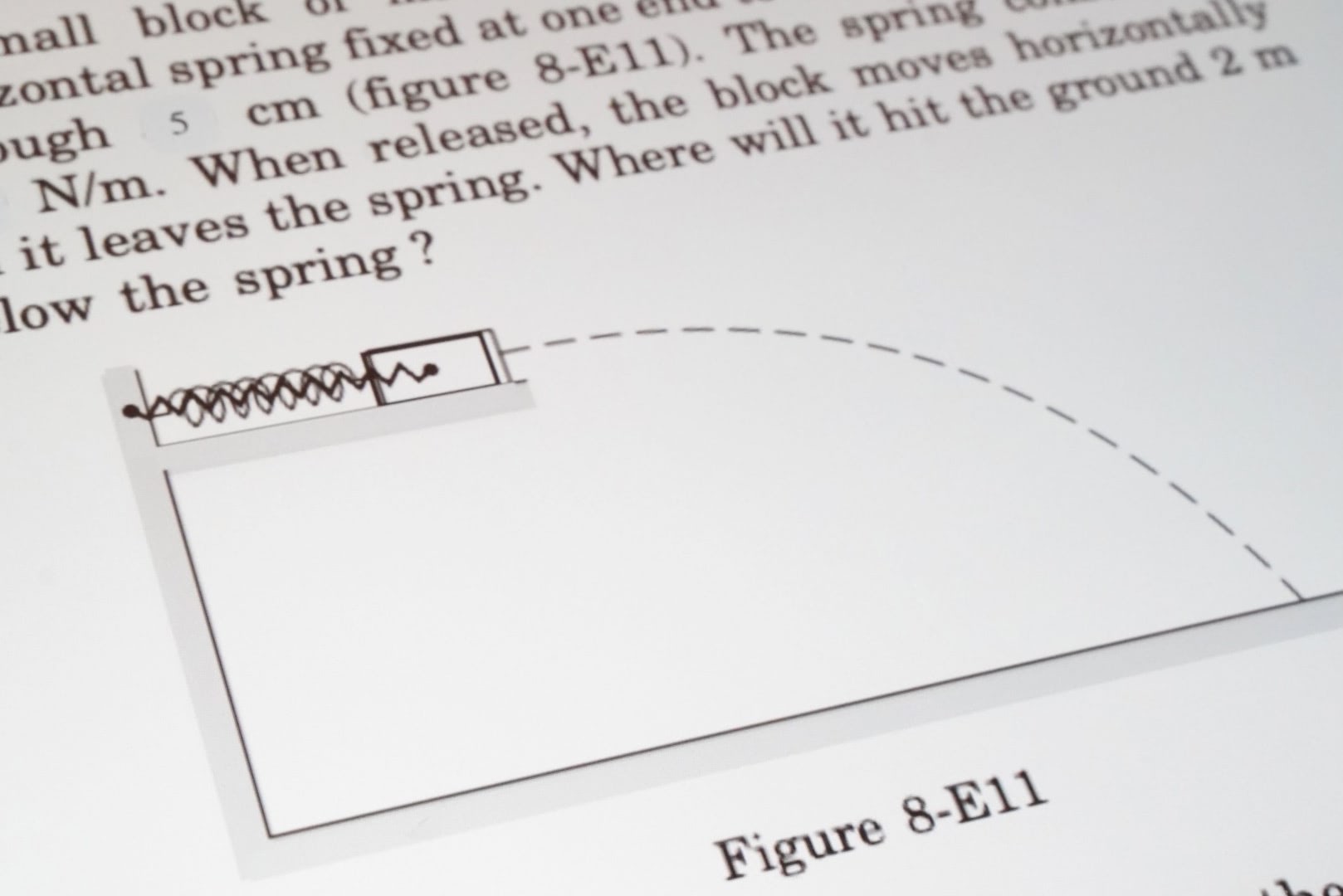}
\includegraphics[width=0.32\linewidth]{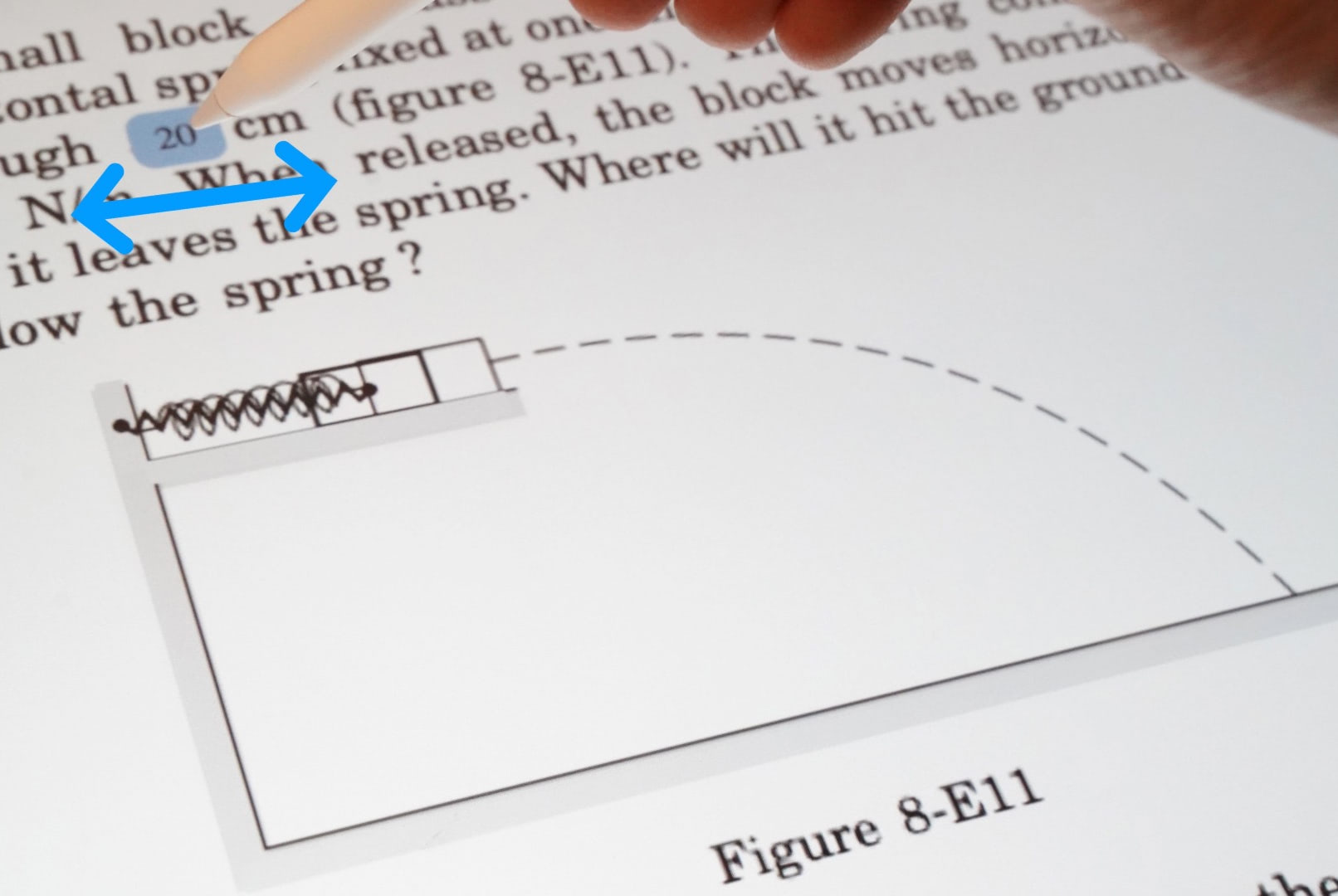}
\includegraphics[width=0.32\linewidth]{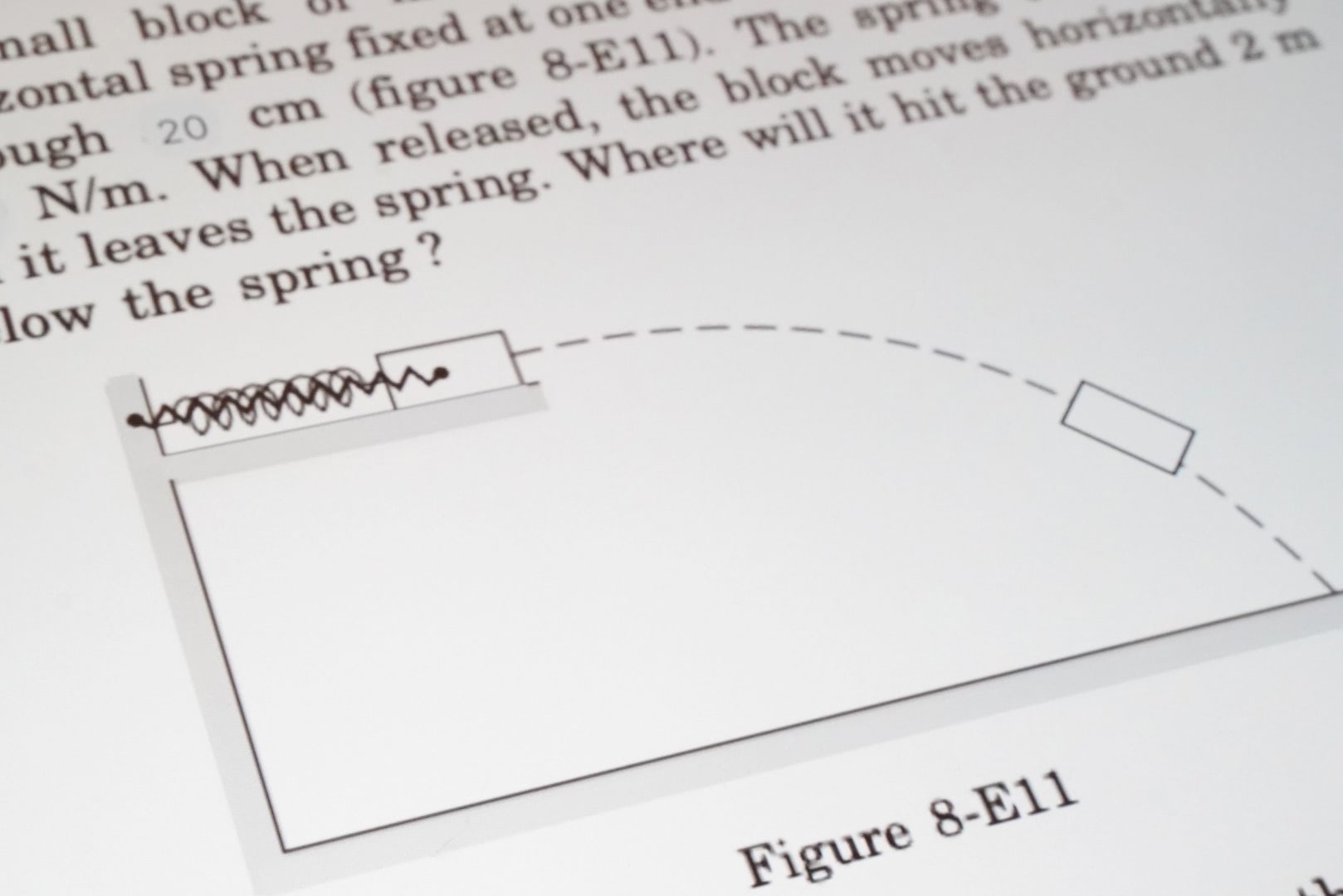}
\caption{bi-directional binding for a kinematics diagram. 1) The user first selects and binds a value. 2) The user drags the value. 3) The system runs the simulation based on the updated value.}
\label{fig:spring1}
\Description{Images of physics textbook being simulated with bi-directional binding.}
\end{figure}

Bi-directional Binding enable authors to link parameter values from the text to the associated simulation. This feature allows students to adjust these values directly within the text and observe the changes in real time. Initially, the system identifies and highlights all numbers within the provided image for the author. Then, the author can select a specific numerical value from the text and assign it a simulation property through a dropdown menu that displays all available properties. For example, Figure~\ref{fig:spring1} illustrates how the user binds the value in the text to the \textit{compression} property, enabling the system to use this value to simulate the scenario by changing the spring's compression.

\subsubsection{\textbf{Parameter Visualization}}
\begin{figure}[h!]
\centering
\includegraphics[width=0.32\linewidth]{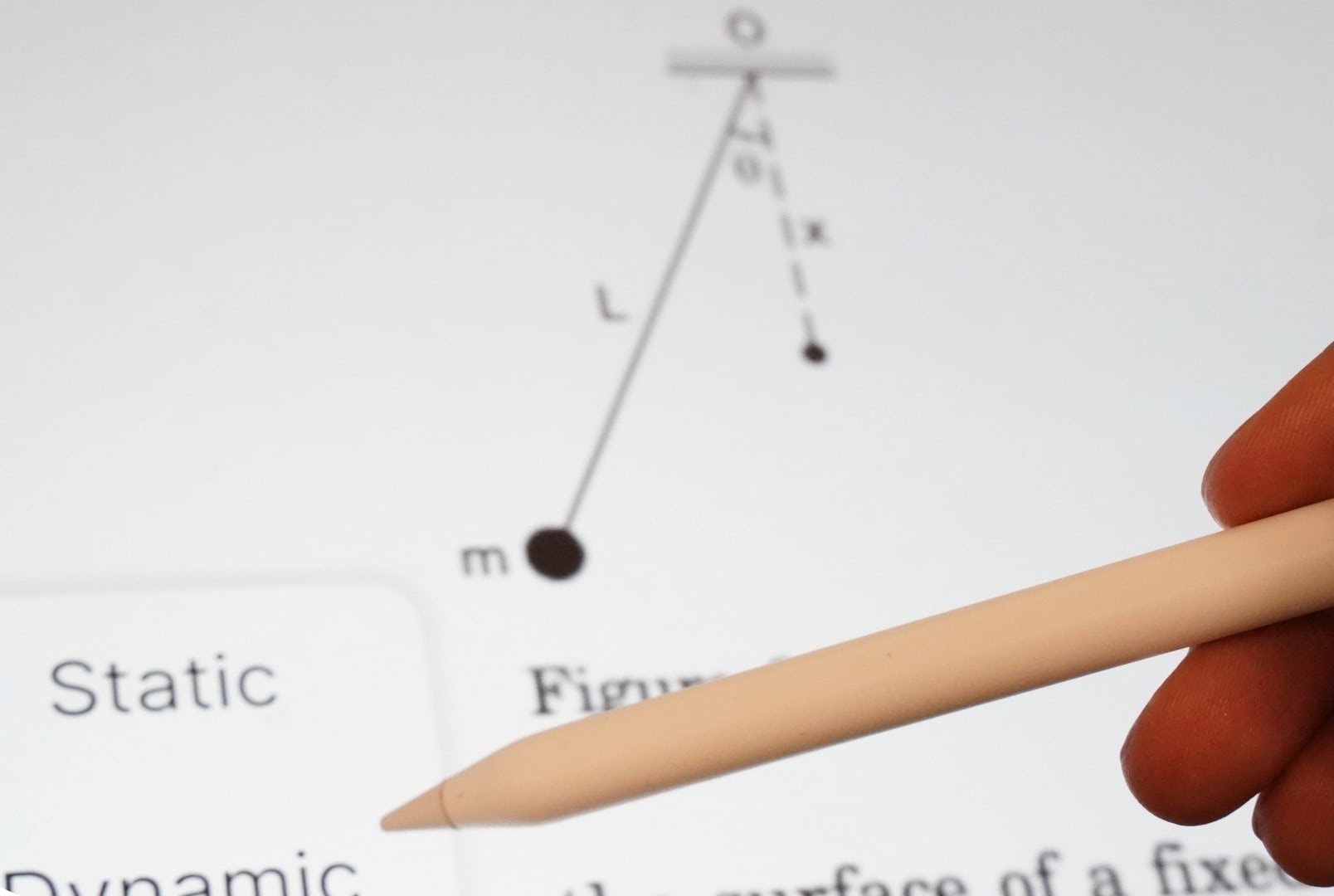}
\includegraphics[width=0.32\linewidth]{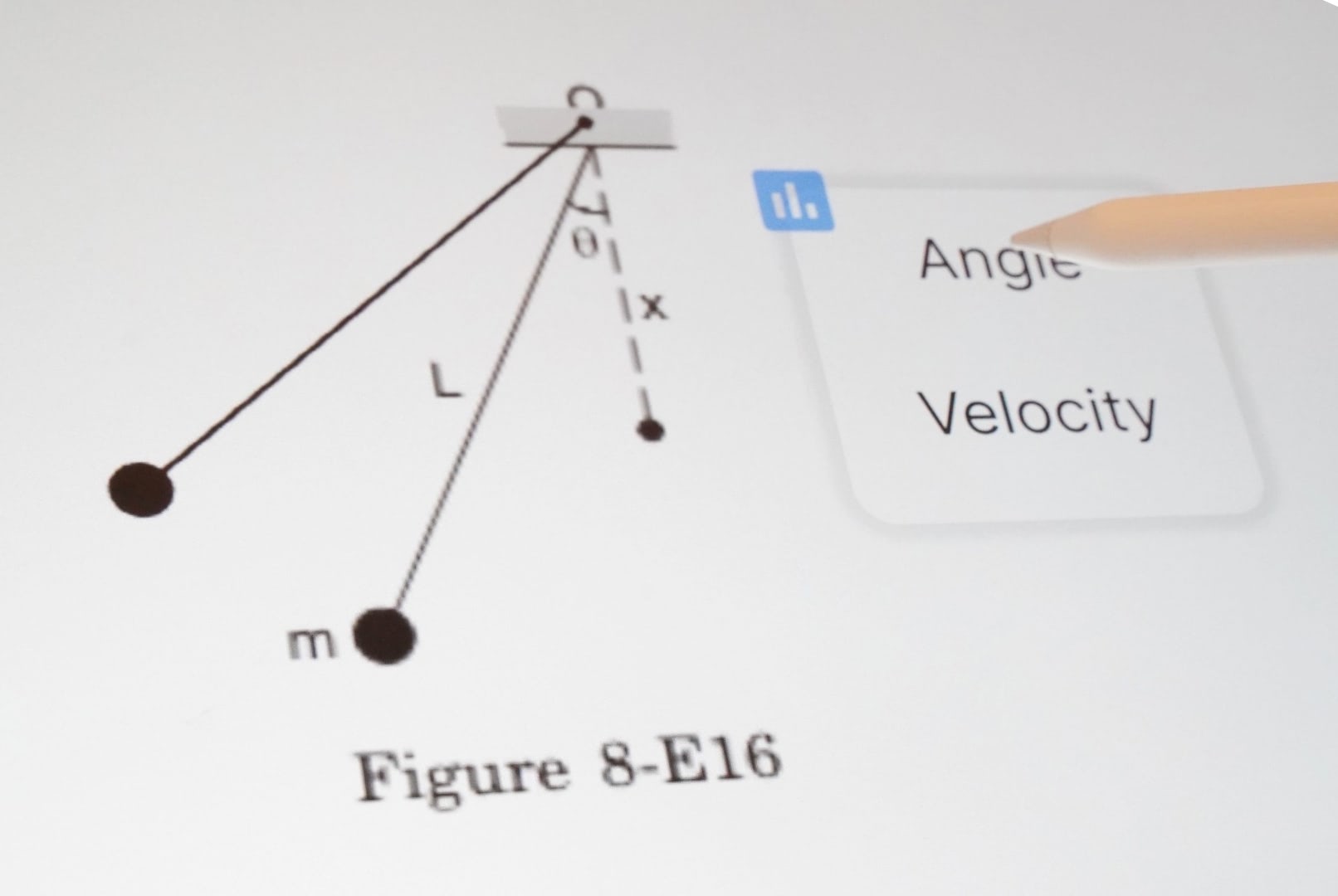}
\includegraphics[width=0.32\linewidth]{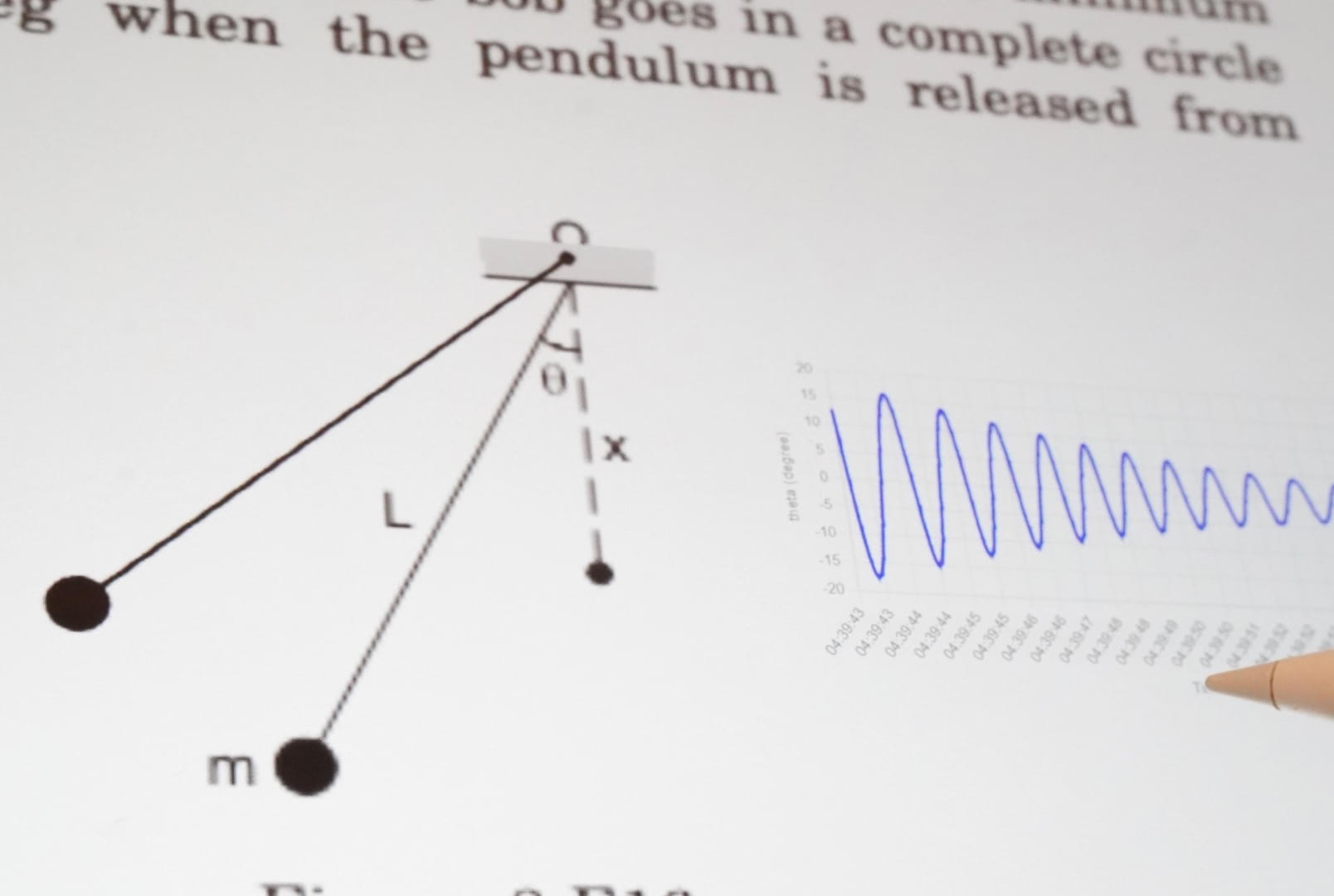}
\caption{Parameter visualization for a pendulum diagram. 1) The user starts simulating a pendulum. 2) The user selects an available parameter. 3) The system visualizes the change over time.}
\label{fig:pendulum1}
\Description{Images of a pendulum being simulated in-situ overlaid on the textbook.}
\end{figure}

Finally, parameter visualization allows authors to visualize the selected value through a dynamic graph. The system visualizes it through a basic time-series graph. For example, in Figure \ref{fig:pendulum1}, a user observes a graph depicting the variation of a pendulum's angle in harmonic motion as it approaches its equilibrium position.

\subsection{Implementation}
Our system comprises two main components: a backend computer vision pipeline module using Python and a frontend web interface developed with React.js. The computer vision module integrates Segment-Anything~\cite{kirillov2023segment}, a widely utilized image segmentation model, alongside custom-developed line and contour detection algorithms through OpenCV. Communication between the frontend and backend is facilitated via the Firebase real-time database, allowing for the processing of images based on provided input coordinates. The results, including extracted images, lines, or points, are subsequently communicated back through Firebase. Moreover, we compute the bounding boxes and X and Y coordinates of extracted image segments and transmit this data to the frontend. For text recognition and the extraction of numerical values, we utilize Google's Cloud Vision API. \adi{We then send the page text and extracted object data in JSON format to an LLM (GPT-4), which is used to recommend simulation type and automatically set the parameters of the simulations based on text.}
For our prototype and technical evaluation, we used Google Chrome on MacBook Air 14 inch 2022 (M2 with 10-Core Integrated GPU and 16GB RAM) for the frontend and Google Colab (CPU: Intel Xeon 4 cores, GPU: Nvidia T4, RAM: 50GB) for the backend.

\subsubsection{Kinematics Simulation}
For our kinematics simulations, we utilize MatterJS\footnote{https://brm.io/matter-js/}, a \adi{popular} JavaScript library for 2D Newtonian physics simulations. The images that users extract are transformed into 2D polygons that reflect their actual shapes. The polygons are subsequently integrated into the physics engine as rigid bodies, with their segmented images acting as sprites. These objects receive user-defined properties, designating them as either static or dynamic bodies, where the former remains stationary and the latter is affected by simulation factors like gravity and time. Additionally, authors have the option to include a spring, a line, or identify an extracted object accordingly.
% authors can designate an extracted image as either a line or a spring. 
The alignment of the segmented images and generated polygons is achieved by matching the bounding boxes of each object, ensuring that rigid bodies are accurately overlaid on the diagram.

\subsubsection{Animated Diagrams}
Our animation pipeline also uses the Segment-Anything model to extract the user-specified path from a diagram, offering a more effective solution than traditional line extraction methods. Utilizing the model's ability to process both positive and negative prompts, users can easily identify their chosen path by clicking on it and marking it as a \textit{path}. This feature allows for the addition of extra path points through further clicks on the image and the exclusion of undesired segments by identifying them as ``negative'' points, thereby improving the precision of path selection and the overall accuracy of path extraction. After isolating the path segment as an image mask, we apply skeletonization and thinning techniques with OpenCV and skimage to refine the mask to a 1-pixel width, effectively removing any extraneous artifacts \adi{and} noise. \adi{The result is a clear set of points defining the intended path. The user can select multiple paths this way and assign them to multiple objects (extracted by SAM) in the diagram.} For the animation execution, we utilize the GSAP.js animation library\footnote{https://gsap.com/}, animating the object along the determined path and integrating additional animation parameters, such as speed and direction, \adi{editable by the user}.

\subsubsection{Optics Simulation}
We developed a custom optics simulator utilizing P5.js visual graphics library\footnote{https://p5js.org}. Our simulator currently supports convex lenses, concave lenses and mirrors. It calculates the positions of two representative light rays based on the object and focal point positions, emulating the common practice of manual diagram drawing. 

\subsubsection{Circuit Simulation}
We developed a custom circuit simulator designed to operate within a web browser, incorporating principles of circuit theory, such as Kirchhoff's laws. Utilizing \adi{the Gemini Multimodal Vision Model (\textit{gemini-1.5-pro})}, our system identifies and segments resistor, capacitor, and battery symbols within circuit diagrams by detecting and extracting bounding boxes. \adi{Contour detection is then applied on the image, which isolates lines and discriminates them based on their orientation.} By identifying junctions within the diagram, the system automatically links the bounding boxes of detected resistors or voltage sources to lines, symbolizing wires. The circuit is represented using a simple array structure, which is transmitted to the web interface through Firebase real-time database and subsequently visualized.

\subsection{Technical Evaluation}
\subsubsection{Method}
We evaluated the accuracy and versatility of our pipeline through technical evaluations. We first gathered six different physics textbooks covering topics such as kinematics, optics, circuit theory, and magnetism. From each textbook, we randomly selected 10 pages containing diagrams relevant to each simulation category (kinematics, circuits, optics, and animation), resulting in a total of 200 diagrams for our sample dataset. We applied our detection pipeline across these diagrams for each simulator category. For object segmentation, we simply select objects \adi{via} mouse interaction. For line segmentation (for animated diagrams), we employ four points, two positive and two negative prompts, to segment the line. After that, multiple authors manually review the results by looking at the generated outcomes due to the absence of a standardized and automated way to check the results, guided by a rubric described below. The complete list of pages and figures evaluated with our system will be provided in the supplementary materials. Our analysis focused on measuring the error rate in various components of the pipeline. 

\subsubsection{Results}
Table~\ref{table:technical-evaluation} presents a summary of our technical evaluation results. The success rates for the different components of the simulation are as follows: kinematics at 64\%, optics at 44\%, circuits at \adi{40\% (62\% with minor edits)}, and animation at 66\%.

Kinematics, Optics, and Animation work through semi-automatic segmentation. Notably, object segmentation demonstrated a high success rate with 86\%. This high success rate for segmentation contributes significantly to the relative success of kinematics, optics, and animation components. Specifically, in kinematics, the success rates for polygon generation and placement are 72\% and 70\%, respectively, indicating effective conversion into physics-simulatable bodies with proper segmentation. However, challenges arise in kinematics simulations due to limitations in supporting certain features (6\%), such as rotational motion, body specific gravity, unsupported objects like ropes, and issues with simulating curved surfaces smoothly. Additionally, we noted that 74\% of the diagrams just required minor adjustments, like modifications to simulation parameters, to achieve accurate simulation results. The success rate without any authoring and modification process was at 40\%. Animation and optics were also consistent with the number, but we observed that the line segmentation success rate was lower, despite using the same Segment Anything technique. For optics in particular, simulation failures comes from diagrams our simulator does not support, such as those with multiple lenses (detecting two as one), prisms, new lens types (like an eye), etc.

% On the other hand, circuit simulations utilize a custom automated recognition model, leading to a notably lower success rate of 15\% for circuits. This lower rate is primarily attributed to the lack of support for various symbols, with 53\% of diagrams containing unsupported elements such as capacitors, inductors, LEDs, and motors. Furthermore, our SVM-based automatic circuit symbol recognition system demonstrated a mere 10\% success rate, hindered by an insufficient training dataset. Our method for circuit connection and line recognition achieved a 45\% success rate. Overall, circuit simulations could perform well given the symbols are correctly recognized. 

\adi{Our Circuit simulation pipeline utilizes a line detection method to localize and identify wires in conjunction with the Gemini model to detect symbols. The line detection success rate is on the lower side at 45\% leading to an overall success rate of simulation at 62\% with minor connection edits. Without any edits the pipeline is successful 40\% of the time. The main reason for this is because of overlapping or crossing wires or artifacts in the diagrams. On the other hand, symbol recognition is reasonably accurate at 72\%. The symbols missed by the Gemini model were mainly due to the error in the bounding boxes returned, which did not align with the symbol.}

\begin{table}[ht]
\centering
\begin{tabularx}{\linewidth}{@{}l X r@{}}
\hline
\multirow{4}{*}{Kinematics} & Object Segmentation & 86\% \\
& Polygon Generation & 72\% \\
& Polygon Placement & 70\% \\
& Simulation & 64\% \\
\hline
\multirow{3}{*}{Animation} & Object Segmentation & 86\% \\
& Line Segmentation & 70\% \\ 
& Animation & 66\% \\ 
\hline
\multirow{2}{*}{Optics} & Object Segmentation & 86\% \\
& Simulation & 44\% \\ 
\hline
\multirow{4}{*}{Circuits} & Line Detection & 45\% \\
& Symbol Recognition & 72\% \\
& Simulation & 40\% \\
& Simulation with Minor Edits & 62\% \\
\hline
\end{tabularx}
\caption{Technical evaluation results}
\label{table:technical-evaluation}
\end{table}

%% file: 5-user-study.tex
\section{User Study}
We evaluated our system through two user studies: 1) a usability study with twelve participants, and 2) expert interviews with twelve physics instructors. The purpose of the usability study was to measure the system's usability and the user's preferences across multiple supported features from a student's perspective. On the other hand, the expert interviews aimed to seek critical feedback on pedagogical aspects from a teacher's perspective.

\subsection{Preliminary User Evaluation}

\subsubsection{Method}
We recruited 12 participants, all of whom possessed at least high school-level knowledge in physics. Each session lasted approximately 40 minutes, and participants were compensated with 15 CAD for their time. After obtaining consent and introducing them to the study, we guided the participants through a demo walkthrough of the system, which included a circuit example (Figure \ref{fig:circuit1}). Once the participants had familiarized themselves with the system, they were allowed to interact with four prepared examples and features, including augmented experiments, animated diagrams, parameter visualization, and bi-directional binding. All participants used the same textbook pages and physics diagrams, ensuring that everyone experienced the same set of features with a standardized set of examples. In our sessions, we utilized two optics examples, one on circuits, one on pendulums, one on kinematics with slope, and one animation featuring a solar system, as described in Section 4. We employed a talk-aloud methodology for the study, encouraging participants to verbalize their feedback as they interacted with the system.

After the experiment, we collected their feedback on the usability of the system. Initially, participants were shown the static diagrams, followed by the animations. We then qualitatively compared their responses to the static diagrams. All sessions were conducted in person. They were also asked to comment on the intuitiveness, engagement, and usefulness of each feature in comparison to static textbooks. Following the talk-aloud session, participants were asked to complete a survey that included questions about the system and its features, as well as a usability questionnaire adapted from the System Usability Scale \cite{brooke1996sus}. 

\subsubsection{Results} 
\begin{figure}[h!]
\centering
\includegraphics[width=\linewidth]{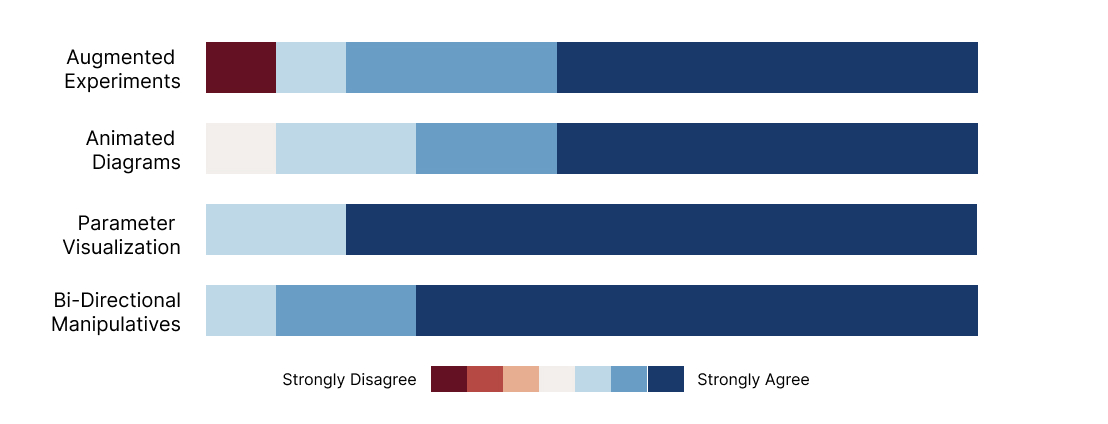}
\caption{Participant feedback on the different types of augmentations for intuition building.}
\label{fig:usability-study}
\Description{A horizontal stacked bar chart of the responses that respondents gave to our user study questionnaire. The length of each colored rectangle corresponds to the number of participants choosing a given response. Dark red denotes “strongly disagree”, orange denotes “disagree”, light orange denotes “somewhat disagree”, white denotes “neutral”, light blue denotes “somewhat agree”, blue denotes “agree” and dark blue denotes “strongly agree”.
Response counts for each question are:
Augmented Experiments help me in intuition building (strongly disagree=1, somewhat agree=1, agree=3, strongly agree=6)
Animated Diagrams help me in intuition building (neutral=1, somewhat agree=2, agree=2, strongly agree=6)
Parameter Visualization helps me in intuition building (agree=2, strongly agree=9)
Bidirectional Manipulatives help me in intuition building (somewhat agree=1, agree=2, strongly agree=8)}
\end{figure}

This section presents the outcomes of our preliminary user study. We evaluated the system usability score (SUS), overall engagement, and the system's usefulness. Our system achieved an overall SUS score of 92.73, with a standard deviation (SD) of 9.84. Participants notably appreciated the Parameter Visualization feature (mean (M)=6.8, SD=0.4) and the Bi-Directional Binding feature (M=6.7/7, SD=0.67) most, followed by the Augmented Experiment (M=6.0/7, SD=1.78) and Animated Diagrams (M=6.2, SD=1.07) features (Figure~\ref{fig:usability-study}). Overall, participants found all of the features useful. 

All participants were asked to talk aloud during all parts of the study. Participants commented on their previous experience studying and understanding physics concepts. One participant mentioned that they \textit{"struggled with understanding abstract concepts (circuits) and that this system made it more intuitive as they can play around with different values and see how it behaves"}. P8 also mentions that \textit{"We can only go to the lab maybe once a week, so for every doubt, I can't do the experiments in the lab to understand it better, so like I feel like we can have more flexibility in experimentation ourselves"}. Participants also found that this tool can help them with confirming or correcting a mental model about how a system functions with respect to the Bi-Directional Binding feature. P7 mentioned that \textit{"Confirming that my understanding of this is what I think it is... Okay, I think I know what's happening but if I play it out and if something doesn't move as expected I can reaffirm my understanding. If I'm doing a problem, like, I want to find where I made the error and I can simply visualize the system with the exact values."}

\subsection{Expert Interviews}
We conducted semi-structured interviews with 12 physics instructors (E1-E12). The goal of the expert interviews was to collect feedback on our tool, evaluate its utility in educational settings, and identify areas for enhancement. These instructors have teaching experience ranging from 1 to 5 years, with an average of 2.4 years. Nine experts teach at the university level, while four have experience teaching at both high school and middle school levels. During the interviews, we allowed them to explore our system and author their own simulations. The interviews lasted approximately 1 hour. The experts were compensated with 35 CAD.

\subsubsection{Method}
The study was conducted through an online meeting on Zoom. First, we inquired about the instructors' backgrounds and their current instructional methods, including the use of external resources like videos or online simulations, the challenges associated with these materials, and their integration strategies within their teaching methodologies. This introductory phase lasted approximately 15 minutes. Subsequently, we introduced the system and allowed them to interact with and create content using a web browser on their computers. Participants were given the option to import and upload their textbooks or diagrams; for those without their materials, we provided a set of ten examples from our technical evaluation dataset (three for kinematics, two for optics, two for circuits, and three for animations). Initially, we demonstrated how to use the system with the first two examples, then allowed the experts to explore and engage with as many examples as they wished. This interactive phase was allocated 15 minutes. Following their engagement with the system, we conducted a semi-structured interview lasting 30 minutes to discuss their experiences. Through open-ended questions, we sought their insights on how our tool compares to existing educational resources and its potential application in their teaching practices.

\subsubsection{Results}
Experts generally conveyed that our system could help them create \adi{personalized} simulations for their students and gave us invaluable feedback about our tool and future implementation. Below we summarize the expert feedback:

% E1 harman E2 uma E3 linan E4 middleschool teacher E5 chen E6 tutorcenter founder and tutor
\subsubsection*{\textbf{Complementing Online Resources, Rather than Replacing}}
Many teachers (E4, E5, E9, and E10) appreciated the contextual and embedded capability, highlighting its unique ability to generate simulations directly from textbook diagrams---a feature that sets it apart from traditional online simulators like OPhysics, which rely on predefined examples. Most experts (E1, E2, E4, E5, E6, E8, E9, E10, E12) incorporate online resources like YouTube videos and simulators in their classrooms, but E4, E6, E8, and E10 noted only 10-20\% of video content aligns with their educational objectives. Providing students with lengthy videos also poses challenges, leading some experts (E9, E10, E12) to use these resources mainly for personal inspiration or to invest considerable time in developing tailored materials.

However, some experts recognize that certain online simulators still offer better pedagogical support due to their ability to facilitate more complex and comprehensive simulations, designed for specific subjects. In this way, both E1 and E6 regard our system as a valuable complement, rather than a replacement, to existing online resources. Despite our relatively simple simulation capabilities, E5 values the system for incorporating an actual physics engine instead of relying solely on animations. E6 similarly appreciates the simplicity and distinctiveness of our simulation features, considering them an effective starting point to help students begin their learning journey. Overall, the participants acknowledged the system's unique contextual support can lower the hurdle to interact with simulations and fill the gap of the current learning tools.

\subsubsection*{\textbf{Extending the Reach of Live Experiments}}
Live experiments and diagrams play a crucial role in physics education, offering tangible insights into theoretical concepts. However, their effectiveness is often constrained by practical limitations. Experts (E3, E4, E5, E6, E12) have highlighted the challenges in preparing live experiments that adequately cover challenging concepts due to lack of resources or time to setup complex experimental equipment. 

\system{} is seen by experts as a promising complement to live experiments, overcoming their limitations through simulations with modifiable parameters and varied scenarios. However, E4 notes that simulations might not capture the \textit{"hands-on learning"} and \textit{"unpredictability of live experiments"}, potentially impacting experiential learning depth. While simulations can enhance comprehension of complex concepts, they may not fully substitute the direct, tactile learning experiences provided by physical experiments, which is seen by experts as a facilitator to deepen understanding of concepts.

% E4, for example, points out the difficulties in demonstrating Newton's first law or varying friction scenarios in class or simulating a perfect condition which is not possible in class. These physical limitations can narrow the range of experiments and lead to student frustration, especially when they cannot execute their creative experiment designs. 

% Moreover, experts (E2, E3, E4) said that accurately drawing complex diagrams in a way that remains clear and understandable to students can become difficult and time-consuming. For example, when they tried to modify states or conditions in a diagram, they had to either erase and modify on top of the existing one or redraw another one with new parameters. They concurred that \system{}, on the other hand, may serve as an effective alternative in these cases. 

% E4, for example, points out the difficulties in demonstrating Newton's first law or varying friction scenarios in class or simulating a perfect condition which is not possible in class. These physical limitations can narrow the range of experiments and lead to student frustration, especially when they cannot execute their creative experiment designs. 

\subsubsection*{\textbf{Engaging Students Through Independent Self-Led Exploration}}
The lack of interest in the subject is a common challenge brought up in teaching physics concepts as explained by E4 and E5. Facilitating genuine interest is a critical component of effective learning. They identify \system{} as a potential tool to ignite the interest of students who initially may not be keen on physics. E2 and E7 commented that \system{} fosters active rather than passive engagement with physics concepts, allowing students to lead their \textit{"own little experiments"}. E4 observes that when students lead their experiments, it enhances independent thought and active learning. This active experimentation could not only cultivate interest but also empower students to delve deeper into physics on their own terms (E4, E5, E6, E9, E10). Experts acknowledged the exploratory potential of the system. Our system may offer an accessible approach for personalized exploration, suggesting a shift towards a more interactive and investigative learning experience in physics.

\subsubsection*{\textbf{Facilitating Questions through Observations}}
Experts (E3, E4, E5, E6, E8, E9, E10) highlighted a prevalent challenge students face in comprehending complex concepts. Educators (E4 and E6) mentioned that our system could aid students in grasping these abstract or difficult-to-visualize concepts. They noted that it allows students to generate more insightful questions based on their observations, thus enhancing their understanding of the concepts. Moreover, they underscored the significance of encouraging students to develop and pose meaningful questions during the learning process, arguing that genuine comprehension stems from meticulous observation of phenomena and subsequent in-depth questioning.

\subsubsection*{\textbf{Need of Verifying Simulations before Classroom Demonstration.}}
When we inquired about the limitations of our system, all experts expressed concerns about its reliability and accuracy. Given its role as a pedagogical tool, teachers would need to verify the simulation results before using them in classrooms. Additionally, they (E4, E5, E6) warned that inaccuracies within the simulations could lead to misconceptions about the concepts being taught, highlighting the critical need for educator oversight to prevent possible misunderstandings. Despite these challenges, all experts still recognized its value in educational settings, provided that they can verify the simulation results beforehand. This verification step could mitigate potential risks and confusion, preventing misunderstandings that may arise from inaccurate results.

% The key, E2 suggests, lies in users' ability to critically \textit{"assess and select the accurate parts of the content"}.
% E2 compared it to other AI technologies, such as ChatGPT, which despite occasional inaccuracies, are still widely used. , and said that no tool or material offers perfect reliability
% Upon encountering inaccuracies within our simulation system during its authoring phase, 
% , given its potential pitfalls. 
% E4 expressed her interest in understanding the system's working mechanism to identify both its strengths and shortcomings accurately. 

\subsubsection*{\textbf{Teachers Emphasize Independent Thought Over Immediate Use of the System}}
Despite its advantages, several experts (E3, E4, E5, E6) emphasize the importance of students engaging with the concept thoughtfully before immediately turning to the simulation. E4 specifically mentions the need for careful timing in introducing such tools to \textit{"avoid disrupting students’ focus in the natural flow of the thought process"}. E6 also shared a rewarding experience of witnessing simulations unfold as his predictions and calculations. He observes that \textit{"such moments not only validate the learning process but also leave a lasting impression on students"}, enhancing their understanding and retention of concepts.

%% file: 6-future-work.tex
\section{Limitations and Future Work}

\subsubsection*{\textbf{\textit{Deploying on a Large Scale}}}
\adi{Despite the support for various physics concepts, the system struggles with complex or abstract illustrations and occasionally fails to detect certain objects. Participants appreciated the diversity of use cases but desired more sophisticated control and broader applicability. Future implementations can include simulators for a wider range of physics topics, like molecular dynamics, and offer more control over the simulations. Experts advocated for more customization, such as control over scenario setup and the ability to add or duplicate custom objects for richer demonstrations while maintaining simplicity to avoid a steep learning curve. E6 and E11 also suggested incorporating visual aids like trajectory paths or frame-by-frame illustrations to address observational challenges, like estimating speeds or remembering sequences. 
A plugin-based system could allow teachers to integrate custom simulators for broader applicability. Moreover, large-scale classroom deployment should be examined, evaluating how instructors and novice learners interact with the system to enhance learning on a larger scale.}

\subsubsection*{\textbf{Integrating with AR Devices}}
Expanding beyond the mobile interface and introducing the system into an augmented reality (AR) headset environment could enhance user engagement by providing a more immersive experience. While current limitations in the precision of \system{} preclude its immediate deployment on AR headsets, future refinements will open avenues for immersive learning experiences beyond mobile and computer-based interfaces.

\subsubsection*{\textbf{AI-Assisted Learning}}
\adi{Although we employ large language models (LLMs) in our pipeline for tasks such as detecting circuit symbols, recommending simulation types, and auto-filling parameter values, there is immense unexplored potential. Previous works have shown great promise in integrating intelligent recommendations into workflows such as online meetings \cite{xia2023crosstalk}, active reading \cite{chen2022marvista}, and research \cite{fok2023scim}. The next step could be intelligent physics tutoring. Integrating multimodal LLMs throughout our pipeline can enhance the authoring process. For instance, these models can automatically detect the context of diagrams, extract static and dynamic parts of images, and create simulations in real-time. Importantly, the system would still allow teachers to edit and reorganize simulations if the LLM makes errors or if the educator prefers to demonstrate a simplified version. Additionally, multimodal LLMs can function as teaching guides or tutoring agents, leading students through concepts step-by-step by creating simpler simulations and progressively increasing complexity. Exploring the efficacy of intelligent physics tutoring can be a promising future direction, integrating our extraction pipeline and augmentation strategies with LLMs in unique ways.}

%% file: 7-conclusion.tex
\section{Conclusion}
We introduced \system{}, a machine learning-integrated tool for transforming static physics diagrams into interactive simulations. 
Using Segment-Anything, OpenCV, and \adi{MLLMs}, our system semi-automatically extracts content from diagrams, enabling non-technical users to create personalized, interactive, and animated explanations without programming. 
Findings from two user studies suggest that our system supports more engaging and personalized learning experiences. Future work includes expanding to broader domains covered in physics, investigating the potential for classroom use through in-the-wild deployments and lastly, exploring the mixed-reality modality to enhance physics education.